\documentclass[prl,aps,superscriptaddress,footinbib,twocolumn]{revtex4-2}
\usepackage[latin9]{inputenc}
\usepackage{color}
\usepackage{amsmath}
\usepackage{mathtools}
\usepackage{amssymb}
\usepackage{stmaryrd}
\usepackage{graphicx}
\usepackage[normalem]{ulem}
\usepackage{bm}
\usepackage{bbm}

\usepackage[unicode=true,bookmarks=true,bookmarksnumbered=false,bookmarksopen=false,breaklinks=false,pdfborder={0 0 1},backref=false,colorlinks=true]{hyperref}

\hypersetup{
 linkcolor=magenta, urlcolor=blue, citecolor=blue, pdfstartview={FitH}, unicode=true}

\makeatletter

\usepackage{amsfonts}\usepackage{tabularx}\usepackage{dcolumn}\usepackage{bm}\usepackage{graphicx}\usepackage{epstopdf}
\usepackage{times}

\usepackage{comment}
\hypersetup{urlcolor=blue}
\definecolor{sec_red}{RGB}{166, 40, 37}

\def\ket#1{\left|#1\right\rangle}
\def\bra#1{\left\langle#1\right|}

\makeatother

\begin{document}

\title{Solving excited states for long-range interacting trapped ions with neural networks}

\author{Yixuan Ma}
\thanks{These authors contributed equally to this work.}
\affiliation{Center for Quantum Information, IIIS, Tsinghua University, Beijing 100084, China}
\affiliation{School of Physics, Xi'an Jiaotong University, Xi'an 710049, China}

\author{Chang Liu}
\thanks{These authors contributed equally to this work.}
\affiliation{Shanghai Qi Zhi Institute,  Shanghai 200232, China}

\author{Weikang Li}
 \email{weikang.lee1999@gmail.com}
\affiliation{Center for Quantum Information, IIIS, Tsinghua University, Beijing 100084, China}

\author{Shun-Yao Zhang}
\affiliation{Lingang Laboratory, Shanghai 200031, China}

\author{L.-M. Duan}
\affiliation{Center for Quantum Information, IIIS, Tsinghua University, Beijing 100084, China}
\affiliation{Hefei National Laboratory, Hefei 230088, China}

\author{Yukai Wu}
\email{wyukai@tsinghua.edu.cn}
\affiliation{Center for Quantum Information, IIIS, Tsinghua University, Beijing 100084, China}
\affiliation{Hefei National Laboratory, Hefei 230088, China}

\author{Dong-Ling Deng}
\email{dldeng@tsinghua.edu.cn}
\affiliation{Center for Quantum Information, IIIS, Tsinghua University, Beijing 100084, China}
\affiliation{Shanghai Qi Zhi Institute,  Shanghai 200232, China}
\affiliation{Hefei National Laboratory, Hefei 230088, China}

\begin{abstract}
\noindent
\textbf{The computation of excited states in strongly interacting quantum many-body systems is of fundamental importance. Yet, it is notoriously challenging due to the exponential scaling of the Hilbert space dimension with the system size. Here, we introduce a neural network-based algorithm that can simultaneously output multiple low-lying excited states of a quantum many-body spin system in an accurate and efficient fashion. This algorithm, dubbed the neural quantum excited-state (NQES) algorithm, requires no explicit orthogonalization of the states and is generally applicable to higher dimensions. We demonstrate, through concrete examples including the Haldane-Shastry model with all-to-all interactions, that the NQES algorithm is capable of efficiently computing multiple excited states and their related observable expectations.
In addition, we apply the NQES algorithm to two classes of long-range interacting trapped-ion systems in a two-dimensional Wigner crystal. For non-decaying all-to-all interactions with alternating signs, our computed low-lying excited states bear spatial correlation patterns similar to those of the ground states, which closely match recent experimental observations that the quasi-adiabatically prepared state accurately reproduces analytical ground-state correlations. For a system of up to 300 ions with power-law decaying antiferromagnetic interactions, we successfully uncover its gap scaling and correlation features. 
Our results establish a scalable and efficient algorithm for computing excited states of interacting quantum many-body systems, which holds potential applications ranging from benchmarking quantum devices to photoisomerization.
}

\end{abstract}

\maketitle

\setlength{\parskip}{4pt}


\begin{figure*}[t!]
  \centering
  \includegraphics[width=1\textwidth]{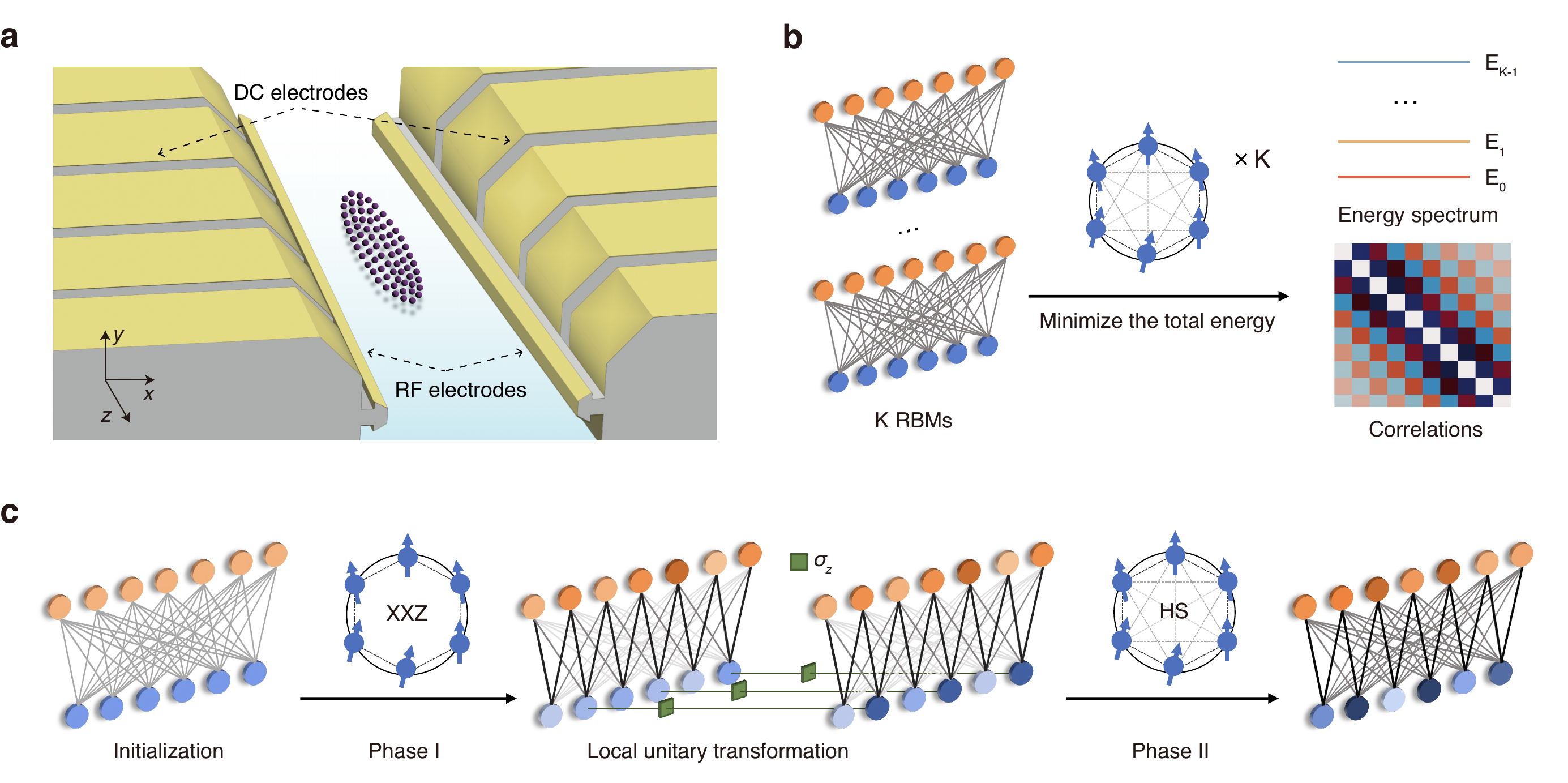}
  \caption{\textbf{Illustration of the trapped-ion platform and the NQES framework.} \textbf{a,} An illustration of the monolithic ion trap where the ion crystal is located on the two-dimensional $xz$ plane. An all-to-all transverse-field Ising model can be realized by shining global laser and microwave pulses to the ion crystal. \textbf{b,} The algorithm to simultaneously learn the first $K$ low-lying eigenstates for a given model. $K$ restricted Boltzmann machines construct a grand function $\Psi$ in a determinant form to ensure their mutual orthogonality. The energy of this grand function is minimized to find the ground state for the composite system, hence the $K$ lowest eigenstates for the original model.  \textbf{c,} The learning procedure for the Haldane-Shastry model as a benchmark. We adapt the curriculum learning framework by progressively increasing the task complexity. 
  In Phase I, we run the NQES algorithm for the XXZ model, and then perform a local unitary transformation to obtain solutions to the antiferromagnetic Heisenberg model.
  In Phase II, we further tailor the Hamiltonian to the desired Haldane-Shastry (HS) model.}
 \label{fig1}
\end{figure*}

Understanding interacting quantum many-body systems is a central task in a wide range of disciplines, from condensed matter physics, quantum chemistry to materials science~\cite{Defenu2023Longrange}. For low-dimensional systems with short-range interactions, tensor-network-based methods have been successful in finding their ground states with area-law entanglement~\cite{White2023Early,Schollwock2005Densitymatrix,Hastings2007Area,Eisert2010Colloquium}. However, for higher-dimensional systems, long-range interactions, or general dynamics involving excited states, classical computational algorithms remain inefficient, whereas quantum simulation has emerged as an appealing approach to investigate their properties~\cite{feynman2018simulating,georgescu2014quantum,cirac2012goals}. Recently, quantum simulators with hundreds of qubits have been demonstrated in various physical systems, including superconducting circuits~\cite{kim2023evidence,King2025Beyondclassical}, neutral atoms~\cite{Ebadi2021Quantum}, and trapped ions~\cite{Bohnet2016Quantum,Guo2024Siteresolved}. In particular, ref.~\cite{King2025Beyondclassical} claims quantum advantage for quench dynamics of a transverse-field Ising model under various topologies of the qubits. Although later tensor-network-based algorithms have been developed to classically reproduce the dynamics in 2D and 3D with nearest-neighbor interactions~\cite{tindall2025dynamicsdisorderedquantumsystems}, the general all-to-all connectivity, as in the case of trapped ions owing to their long-range Coulomb interaction, still remains a notable challenge. 
Such difficulty complicates experimental validation. For example, ref.~\cite{Guo2024Siteresolved} verifies a 300-ion quantum simulator by engineering a special class of Ising Hamiltonian whose ground state can be exactly solved, and by comparing the measured spin correlations in the quasi-adiabatically prepared ground state with the theoretical predictions. Surprisingly tight agreement has been obtained even though the experimental ramp time, limited by the spin coherence time of the trapped ions, is much shorter than the adiabatic time dictated by the vanishing energy gap near the quantum critical point. Since truncated ramps inevitably populate excited states, the observation poses a puzzle: why do the measured correlations still mimic those of the ideal ground state? Therefore, a theoretical understanding of such excited states will be highly desirable to fully interpret the current quantum simulation results and to better validate the quantum simulators.

Yet, computing excited states of a quantum many-body system is a challenging task, especially when the interactions are long-range. 
Recent years have witnessed the growing success of neural network representations for approximating complex wavefunctions, particularly for ground-state problems. Architectures such as variational autoregressive models~\cite{Sharir2020Deep,Barrett2022Autoregressive}, restricted Boltzmann machines~\cite {Carleo2017Solving,Deng2017Quantum,Carrasquilla2021How,Nomura2021Helping}, and fermionic neural networks~\cite{Hermann2023Initio} have demonstrated intriguing expressivity and scalability across a variety of settings, from lattice spin models to electronic structure calculations. These advances have sparked growing interest in extending neural network representations to excited-state problems. However, existing neural-network approaches for excited states remain limited, with the explicit necessity of enforcing orthogonality: they typically incorporate orthogonality constraints by penalizing wavefunction overlap during optimization~\cite{Li2024Spinsymmetryenforced,Entwistle2023Electronic,Wheeler2024Ensemble}.
This procedure requires delicate hyperparameter tuning and may bias optimization. Alternatives based on stochastic generalized eigenvalue problems offer partial relief~\cite{Schautz2004Optimizeda,Cordova2007Troubleshooting,Filippi2009Absorption,Cuzzocrea2020Variational,Dash2021Tailoring} but are restricted to linear ans\"atze and are incompatible with expressive neural network architectures~\cite{Ceperley1988Calculation,Nightingale2001Optimization}. Thus, a general-purpose framework for many-body spin systems that can compute multiple excited states simultaneously, robustly, and at scale\textemdash especially in the presence of long-range correlations and sign-structured wavefunctions\textemdash has yet to be realized.

Here, we introduce the neural quantum excited-state (NQES) algorithm: a scalable variational framework for simultaneously computing multiple low-lying excited states for many-body quantum spin systems (Fig.~\ref{fig1}). Our approach is conceptually inspired by the recent framework introduced in ref.~\cite{Pfau2024Accurate} for computing electronic excitations of molecular systems by solving Schr{\"{o}}dinger equation in continuous real space. We extend this paradigm to many-body spin systems, which feature long-range and frustrated interactions.
At its core, NQES employs a determinant-based construction, where $K$ independently parameterized restricted Boltzmann machines are combined into a Slater-determinant-like function for variation. This composite ans\"atz inherently enforces orthogonality among the $K$ components, thereby eliminating the necessity for explicit projections or penalty terms during optimization. To ensure scalability, we develop a memory-efficient stochastic reconfiguration scheme, leveraging Krylov subspace solvers that bypass the need to store or invert large covariance matrices, which renders the optimization of networks with millions of parameters practical. This is further accelerated by a compact spin encoding that translates Pauli operations into bitwise instructions, enabling efficient Monte Carlo updates on classical hardware. To tackle the challenges posed by long-range entanglement and sign problems\textemdash especially in models such as the Haldane-Shastry chain\textemdash we design a \emph{physics-informed curriculum learning} pipeline, which guides the network to gradually acquire the low-lying eigenstate properties.

We benchmark the NQES algorithm across multiple paradigmatic models\textemdash including the transverse-field Ising chain and the long-range Haldane-Shastry model\textemdash and further validate its applicability to experimental trapped-ion systems. For systems with up to 128 spins, NQES accurately recovers the first several excited states of the Haldane-Shastry chain, achieving sub-$10^{-3}$ relative energy errors and faithfully reproducing long-range spin correlations. 
We further apply the NQES algorithm to study two classes of strongly interacting trapped-ion systems arranged in a two-dimensional Wigner crystal. The first model features non-decaying all-to-all Ising interactions with alternating signs, corresponding to the setup of a recent experiment~\cite{Guo2024Siteresolved}. With the same interaction parameters, we compute low-lying excited states under external magnetic fields applied during quasi-adiabatic ground-state preparation. Remarkably, these excited states exhibit spatial correlation patterns closely resembling those of the ground state, both in the paramagnetic and ferromagnetic regimes, which offer insight into why experiments reproduce ideal correlations despite nonadiabaticity. In parallel, the second model is a long-range antiferromagnetic Ising system with power-law decaying interactions. For this case, we study gap scaling across varying field strengths and system sizes, and characterize their excited-state correlations with up to 300 ions. These results demonstrate the expressivity and scalability of our approach in regimes inaccessible to exact diagonalization or tensor network methods. More broadly, the NQES framework provides a versatile computational tool for probing excitation spectra in quantum matter, benchmarking quantum simulators, and modeling photochemical and nonadiabatic processes in strongly correlated systems.

\vspace{.3cm}
\noindent\textbf{\color{sec_red}\large{}The NQES algorithm}

Our framework begins with constructing an expanded function from single wavefunctions, where the $K$ lowest eigenstates can be learned simultaneously with their mutual orthogonality automatically ensured~\cite{Pfau2024Accurate}. 
Consider $K$ independently parametrized wavefunctions $\psi_1({S}),\ldots,\psi_K({S})$, where $S=\{s_1,\dots,s_N\}$ denotes a spin configuration of an $N$-body system. 
To construct a composite function for $K$ orthogonal states, we define an expanded function in the following form:
\begin{equation}\label{eq:GrandWaveMatrix}
{\Psi}(\mathcal{S}) = \mathrm{det}\left[\begin{array}{ccc}
\psi_1({S}^1) & \cdots & \psi_K({S}^1) \\
\vdots & & \vdots \\
\psi_1({S}^K) & \cdots & \psi_K({S}^K)
\end{array}\right],
\end{equation}
where $\mathcal S = (S^1,\ldots,S^K)$ is the collective spin configuration for the function $\Psi$. 
The determinant ensures antisymmetry: $\Psi$ vanishes if any two single states $\psi_i({S})$ and $\psi_j({S})$ become linearly dependent, thereby naturally preventing spectral collapse during optimization.
Then we define a function $|\Psi\rangle=\sum_\mathcal{S}\Psi(\mathcal{S})|\mathcal{S}\rangle$ and utilize it as a variational ans\"atz for a composite system  ${\mathcal{H}}=\sum_{i=1}^K{H_i}$, where $H_i$ acts on the $i$th configuration subspace. These single states will remain linearly independent during the training process. With the total energy $\langle \Psi|\mathcal H|\Psi\rangle/\langle \Psi|\Psi\rangle$ minimized, it will yield the sum of the $K$ lowest eigenenergies for a single system. 
Meanwhile, the obtained single states will be linear combinations of the $K$ lowest eigenstates. By post-processing the trained neural-network functions, we can efficiently evaluate the total energy and compute observable expectation values for each excited state (Supplementary Section~II).

Although our approach is generally applicable to any variational wavefunctions, in this work we focus on the representation by restricted Boltzmann machines (Fig.~\ref{fig1}\textbf{b}, Supplementary Section~I). This ans\"atz defines a wavefunction via
\begin{equation}
    \psi_W(S)=\sum_{\{h_j\}}e^{\sum_{i} a_i s_i+\sum_{j} b_j h_j+\sum_{i j} w_{i j} s_i h_j}, 
\end{equation}
where $W = \{a_i, b_j, w_{ij}\}$ are complex-valued variational network parameters~\cite{Carleo2017Solving,Deng2017Quantum,Carrasquilla2021How}. In the limit of large hidden-layer widths, it is a universal function approximator and can efficiently encode a broad class of entangled quantum states. 
In contrast to tensor-network approaches such as the density matrix renormalization group~\cite{White2023Early}, projected entangled pair states~\cite{Verstraete2008Matrix}, or multiscale entanglement renormalization ans\"atz~\cite{Cincio2008Multiscale}, which are efficient primarily for area-law states, our framework imposes no intrinsic constraint on the entanglement structure. It can compactly represent volume-law entangled wavefunctions and remains applicable across different spatial dimensions. This flexibility makes them particularly suitable for solving excited states in high dimensions with massive entanglement.

\vspace{.3cm}
\noindent\textbf{\color{sec_red}\large{}Performance benchmarks}
 \begin{figure*}[t!]
  \centering
  \includegraphics[width=1\textwidth]{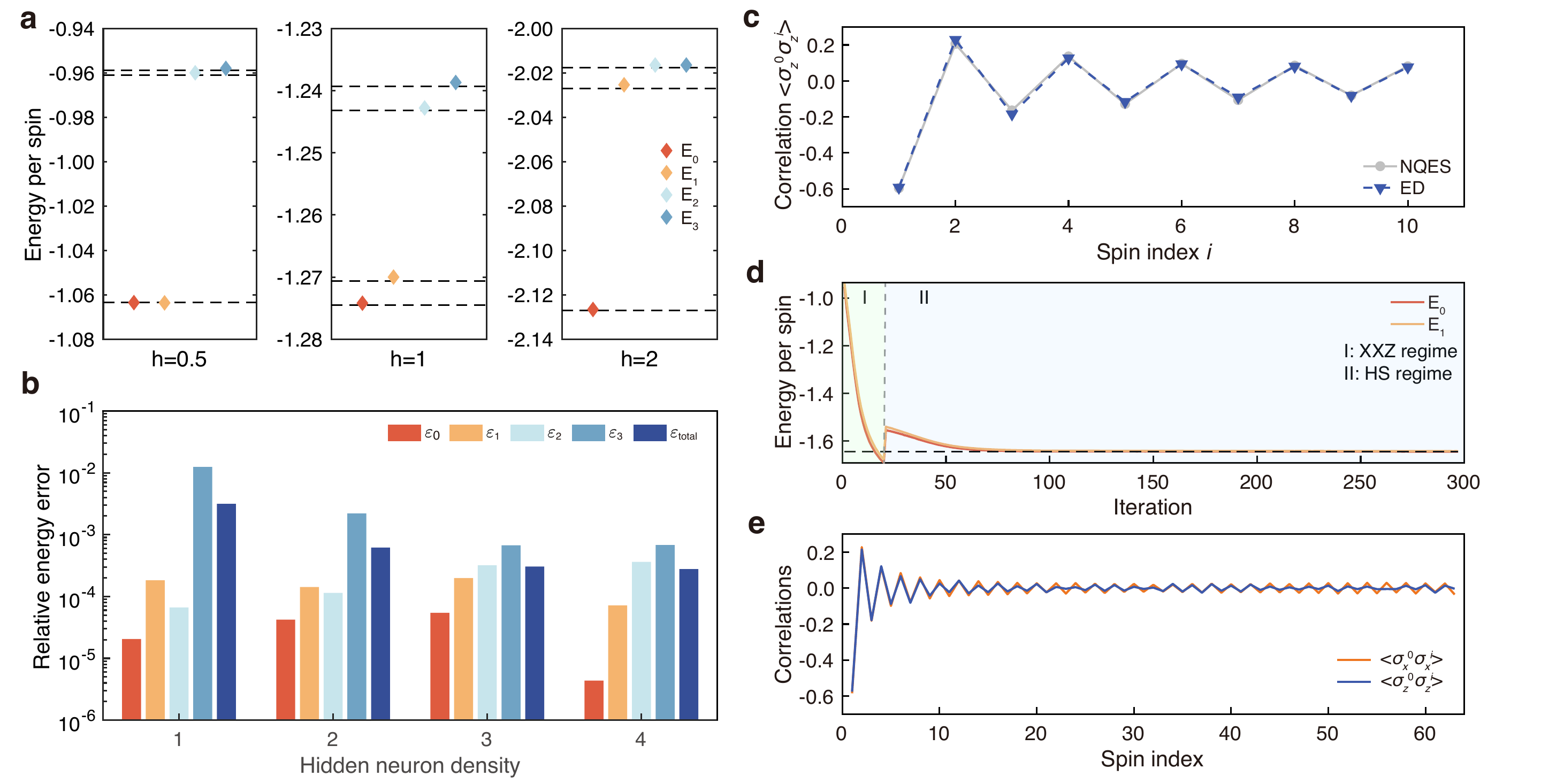}
  \caption{\textbf{Performance benchmarks for the NQES algorithm. }\textbf{a,} Low-lying energy spectra for the one-dimensional transverse-field Ising model with a nearest-neighbor interaction between $N=20$ spins under different transverse fields $h$. Diamonds represent the energies computed with the NQES algorithm, while the dashed lines show the exact diagonalization (ED) results. \textbf{b,} Relative errors for the energy computed with different numbers of hidden neurons $M$, where $\epsilon_i$ ($i=0,1,2,3$) denotes the relative errors of the four lowest eigenenergies respectively and $\epsilon_{\mathrm{total}}$ denotes the relative error for total energy evaluation. Here $h=1.5$. The errors decrease as the hidden neuron density $M/N$ increases. \textbf{c,} Correlation functions $\langle \sigma^1_z\sigma^{i+1}_z\rangle$ ($i=1,\cdots,10$) for the ground state of the Haldane-Shastry model with $N=20$ spins. The correlations obtained from the NQES algorithm (gray dots) agree well with those computed from ED (blue triangles). \textbf{d,} Optimization process for energies of the Haldane-Shastry model with 128 spins. We employ a curriculum learning approach, initializing training on the XXZ model as Phase I up to the point marked by the vertical dashed line. After a local unitary transformation, in Phase II, the system is switched to the Haldane-Shastry model. As training progresses, the predicted energies converge toward the exact eigenvalues (horizontal dashed line). \textbf{e,} Correlation functions $\langle \sigma^1_z\sigma^{i+1}_z\rangle$ and $\langle \sigma^1_x\sigma^{i+1}_x\rangle$ ($i=1,\cdots,63$) for the ground state of the Haldane-Shastry model with $N=128$ spins.}
 \label{fig2}
\end{figure*}

To benchmark the robustness and accuracy of our algorithm, we apply it to two prototypical spin models that are exactly solvable. The first one concerns the transverse-field Ising model with periodic boundary conditions:
${H_{\mathrm{Ising}}} = -\sum_{i} \sigma^i_z \sigma^{i+1}_z + h\sum_i \sigma^i_x$, 
where $\sigma_x$, $\sigma_z$ are Pauli matrices.
We apply the NQES algorithm to a 20-spin Ising chain and compare the results with exact diagonalization (ED). Fig.~\ref{fig2}\textbf{a} shows the lowest four eigenenergies obtained by NQES alongside the ED benchmarks; the two sets of values agree well for varying $h$.  To quantify the accuracy, we define the relative error
$\epsilon=\lvert E-E_{\mathrm{ED}}\rvert/\lvert E_{\mathrm{ED}}\rvert$. In Fig.~\ref{fig2}\textbf{b}, we plot $\epsilon$ for each of the four states, as well as for the total energy, versus the hidden neuron density which is defined as $M/N$. The accuracy improves systematically with increasing network size: once the number of hidden neurons reaches $M=4N$, all individual-state errors fall below $10^{-3}$.

The second model we consider is the Haldane-Shastry model. Compared with the above Ising model with only local interactions in a one-dimensional chain, the Haldane-Shastry model is a paradigmatic spin system featuring all-to-all antiferromagnetic interactions~\cite{Haldane1988Exact,Shastry1988Exact}.
Since it has an exactly solvable energy spectrum, we exploit this model to further assess and benchmark the capability of our method in capturing long-range quantum correlations.
The Hamiltonian of the Haldane-Shastry  model is given by
\begin{equation}
H_1=\sum_{i<j} \frac{1}{d_{ij}^2}\left({\sigma}^i_x {\sigma}^j_x+{\sigma}^i_y {\sigma}^j_y+{\sigma}^i_z{\sigma}^j_z\right),
\end{equation}
where the distance $d_{ij}=(N / \pi)|\sin [\pi(i-j) / N]|$ corresponds to the chord length between lattice sites uniformly distributed on a ring (Supplementary Section~III). The coupling strength between spins decays quadratically with their distance.

Empirically, we observe that training on the Haldane-Shastry model with randomly initialized NQES does not converge. Upon in-depth analysis, we determine that this instability arises from an intrinsic sign problem (Supplementary Section~III~B): the long-range interactions induce complex, nontrivial wavefunction phase structures that cannot be removed by any local unitary transformation~\cite{Bravyi2023Rapidly}. During the optimization procedure, these rapidly fluctuating signs lead to severe cancellations, which in turn cause large statistical variances in energy estimation. In practice, this leads to premature stagnation in optimization, as the parameter updates are dominated by noise well before convergence is reached\textemdash a behavior consistent with our empirical observations.

To address the challenges posed by long-range entanglement and the sign problem, we design a curriculum learning~\cite{Bengio2009Curriculum} strategy that incrementally increases the complexity of the training task, as illustrated in Fig.~\ref{fig1}\textbf{c}. This procedure begins with training on the XXZ model ${H}_{\mathrm{XXZ}}=  \sum_{i}\left(-{\sigma}_x^i {\sigma}_x^{i+1}-{\sigma}_y^i {\sigma}_y^{i+1}+{\sigma}_z^i {\sigma}_z^{i+1}\right)$ from random initialization. Once this is done, we apply a local unitary transformation $U$ to map the learned states to eigenstates of the antiferromagnetic Heisenberg model ${H}^{\prime}= U^\dagger {H}_{\rm XXZ} U = \sum_{i}\left({\sigma}_x^i {\sigma}_x^{i+1}+{\sigma}_y^i {\sigma}_y^{i+1}+{\sigma}_z^i {\sigma}_z^{i+1}\right)$. These eigenstates already contain the staggered sign pattern that dominates the Haldane-Shastry spectrum, giving a physically informed set of initial parameters. Starting from this antiferromagnetic-Heisenberg baseline, we switch on the full Haldane-Shastry interaction and continue optimization, enabling the network to refine its structure and capture the intricate long-range correlations progressively.

\begin{figure*}[t!]
    \centering
    \includegraphics[width=1\textwidth]{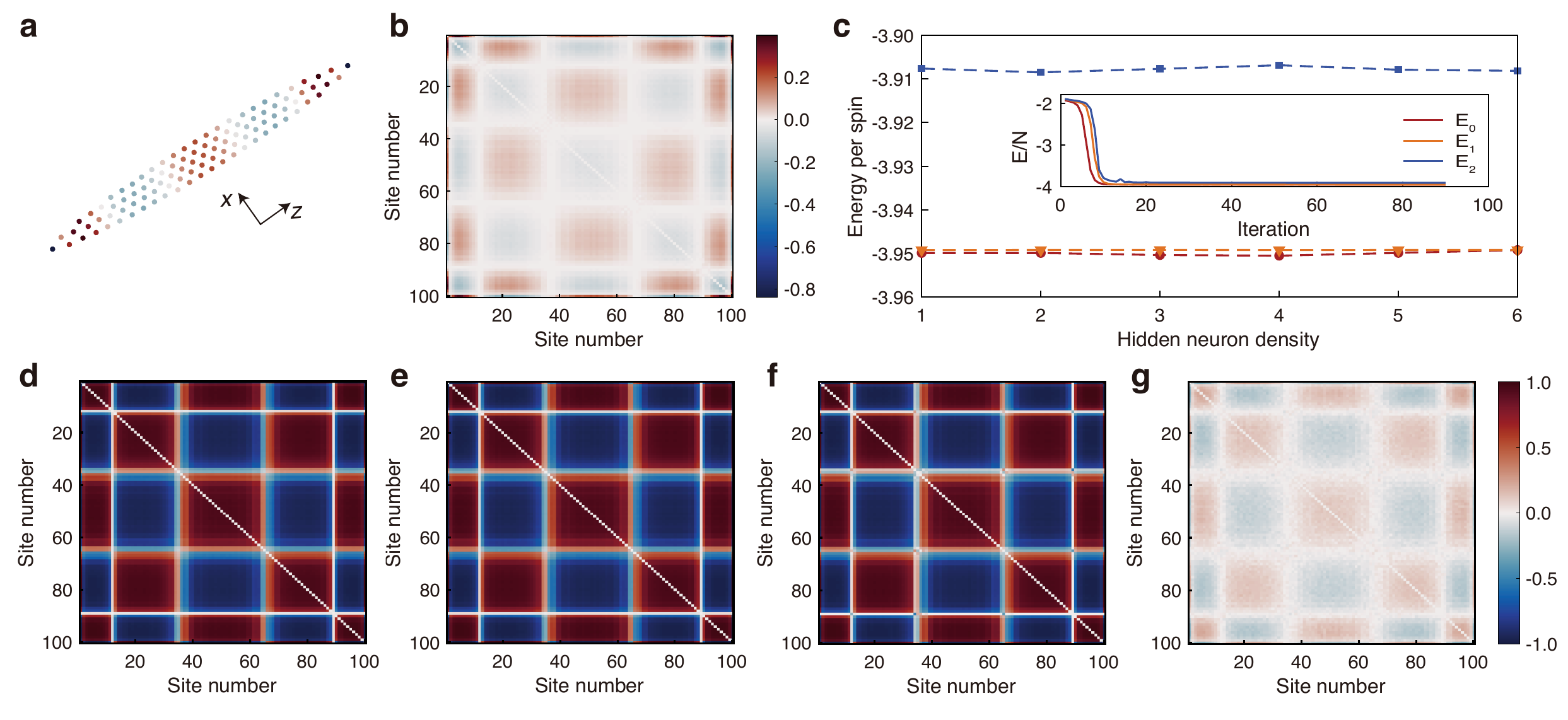}
    \caption{\textbf{Applications to a long-range Ising model mediated by a collective phonon mode of a 2D ion crystal.}  \textbf{a,} Equilibrium positions of $N=100$ ions. Colors represent the amplitudes of individual ions in the seventh highest drumhead phonon mode: red for positive and blue for negative values. \textbf{b,} Theoretically computed Ising coupling coefficients $J_{ij}$ by adiabatic elimination of the collective phonon modes. \textbf{c,} Learning results of the NQES algorithm for a weak transverse field $h=2$ in the ferromagnetic phase. Computed energies for the three lowest eigenstates converge well as the number of hidden neurons $M$ increases. The inset shows the training process. \textbf{d-f,} Two-spin correlations $\left\langle\sigma_z^i \sigma_z^j\right\rangle$ ($i,j=1,2,...,100$) for the lowest three eigenstates, respectively, with $h=2$ and the number of hidden neurons $M=3N$. $\textbf{g},$  Two-spin correlations $\left\langle\sigma_z^i \sigma_z^j\right\rangle$ ($i,j=1,2,...,100$) for the second excited state for a strong transverse field $h=8$ in the paramagnetic phase.}
    \label{fig3}
\end{figure*}

The NQES algorithm equipped with curriculum learning significantly enhances the convergence of the learning procedure. 
We present results for the Haldane-Shastry model in Fig.~\ref{fig2}\textbf{c-e}. In Fig.~\ref{fig2}\textbf{c}, we compare the ground-state correlation function $\left\langle\sigma_z^1 \sigma_{z}^{i+1}\right\rangle,i=1,...,10$ computed by our method with exact diagonalization for a system of 20 spins. The excellent agreement demonstrates that our algorithm not only achieves accurate energy estimates but also faithfully captures long-range spin correlations.
Fig.~\ref{fig2}\textbf{d} illustrates the training process of the ground state and the first excited state for a 128-spin system. Following our curriculum learning strategy, we evaluate the energy on the XXZ model during phase I, then on the Haldane-Shastry model in phase II. There is an increase in energy when we switch the training model, but it converges quickly afterwards. The resulting energy estimates for both ground and excited states agree with the exact solutions within a relative error of $10^{-3}$.
In Fig.~\ref{fig2}\textbf{e}, we show the ground-state correlations along the $x$ and $z$ directions for the 128-spin Haldane-Shastry  model, revealing that the symmetry and long-range structure are accurately learned.

\vspace{.3cm}
\noindent\textbf{\color{sec_red}\large{}Trapped ions}

Next, we apply the NQES algorithm to long-range Hamiltonians that are relevant to current large-scale ion trap quantum simulators (Fig.~\ref{fig3}). Recently, 300 ionic qubits have been realized in a 2D Wigner crystal, with a laser-induced long-range Ising coupling mediated by the collective spatial oscillation modes of the ions~\cite{Guo2024Siteresolved}. In general, we have all-to-all Ising interactions between all the ion pairs:
\begin{equation}
 H_z= \sum_{i\ne j} J_{ij} \sigma^i_z \sigma^j_z,
\end{equation}
where the coupling strength $J_{ij}$ between ion $i$ and ion $j$ can be tuned by the laser frequencies and amplitudes (Supplementary Section~IV). Combined with a transverse field $H_x=-h\sum_{i}\sigma_{x}^{i}$ which can be realized by a global laser or microwave, the eigenstates and dynamics of the system are challenging for classical simulation~\cite{King2025Beyondclassical,Guo2024Siteresolved}.

Specifically, we examine two types of Ising interactions that are relevant to the experiments. First, we consider an all-to-all Ising coupling generated by tuning the laser frequency close to a single drumhead phonon mode of the 2D ion crystal, as plotted in Fig.~\ref{fig3}\textbf{b}. 
In this scenario, a two-fold degenerate ground state can be predicted at zero transverse field ($h=0$) which imprints the structure of the coupled collective phonon mode, and the spin-spin correlation can be computed as $\langle \sigma_z^i \sigma_z^j \rangle = -\mathrm{sign}(J_{ij})$ (Supplementary Section IV~C). Qualitatively, this spatial pattern has been observed experimentally by initializing the system in the ground state $|+\rangle^{\otimes N}$ of $H_x$ and slowly tuning down the transverse field to obtain the final Hamiltonian $H_z$~\cite{Guo2024Siteresolved}. However, the good agreement between the ideal and the experimentally observed patterns remains an intriguing puzzle due to the inevitable nonadiabatic excitations to higher energy levels during the experimental quench time, which is much shorter than the timescale corresponding to the vanishing energy gap at the critical point. 

\begin{figure*}[t!]
    \centering
    \includegraphics[width=1\textwidth]{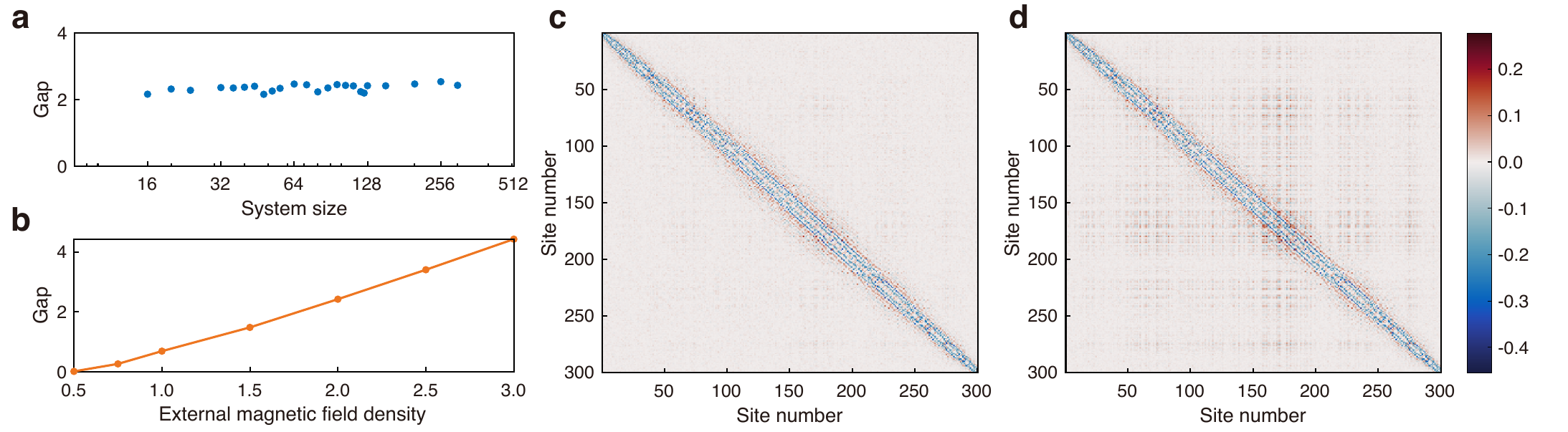}
    \caption{\textbf{Applications to a power-law decaying antiferromagnetic Ising model in 2D ion crystals.} \textbf{a,} Energy gaps for different system sizes with $h=2$. The energy gap is above a constant value independent of the system size. \textbf{b,}
      Energy gaps as a function of transverse field strength for 104 ions. The gap increases as the strength of the external field increases.
      \textbf{c-d,} Correlation maps $\left\langle\sigma^i_z \sigma^j_z\right\rangle,i,j=1,2,...,300$ of the computed ground state and the first excited state for 300 ions with $h=1.2 $ and $M=3N$.} \label{fig4}
\end{figure*}

To understand the robustness of the spatial correlation pattern against nonadiabatic excitations, we apply the NQES algorithm to investigate the low-lying eigenstates in the finite-field regime ($h \neq 0$) as shown in Fig.~\ref{fig3}\textbf{c-g}. By systematically scaling up our neural network model, we observe that energy estimates for the three lowest eigenstates improve in accuracy as the number of hidden neurons $M$ increases from $N$ to $3N$, achieving energy reductions on the order of 0.1\%.
Interestingly, further increasing $M$ to $5N$ slightly reduces accuracy, which might be caused by the difficulties in optimizing a highly complex parameter landscape. Consequently, we identify $M=3N$ or $4N$ as optimal, balancing computational efficiency with representational power (Fig.~\ref{fig3}\textbf{c}).
Fig.~\ref{fig3}\textbf{d-f} presents the spin-spin correlation maps for the three lowest eigenstates with $h$ set to 2 in the ferromagnetic phase.  Although the transverse field is finite, the ground-state correlations (Fig.~\ref{fig3}\textbf{d}) still resemble the phonon-mode pattern obtained in the zero-field limit.  Surprisingly, the first and second excited states exhibit correlation profiles nearly indistinguishable from those of the ground state.  We also compute the excited states for a strong transverse field $h=8$ in the paramagnetic phase. 
We present the correlation map of the second excited state in Fig.~\ref{fig3}\textbf{g}, with the ground-state and first-excited-state correlations provided in Supplementary Section IV. The eigenstates exhibit similar correlation patterns in this regime. These results provide theoretical support that, even in the absence of precise adiabatic control, nonadiabatic excitations have a minimal impact on the spatial correlation patterns observed in experiments.

The second scenario involves a long-range antiferromagnetic coupling following approximately a power law with the ion-ion distance $J_{ij}\propto 1/|\boldsymbol{r}_i-\boldsymbol{r}_j|^\alpha$. This can be generated when the laser frequency is detuned further away from all the phonon modes~\cite{Monroe2021Programmable}. Here we set $\alpha=1$ and rescale $J_{ij}$  so that the nearest-neighbor coupling strength is 1.0. 
We compute the energy gap between the ground state and the first excited state for system sizes ranging from 16 to 300 ions at $h=2$ (Fig.~\ref{fig4}\textbf{a}). The gap remains finite and approximately size-independent, signaling a gapped phase. We also track the gap as a function of $h$ for a fixed chain of 104 ions (Fig.~\ref{fig4}\textbf{b}). The gap decreases simultaneously with $h$, and closes at around $h=0.5$. We then analyze a 300-ion system at $h=1.2$. The correlation functions of both the ground and first excited states are presented in Fig.~\ref{fig4}\textbf{c-d}. Notably, while the ground state exhibits short-range order, the first excited state reveals a more intricate pattern of long-range correlations: The ground-state correlation pattern is not robust against nonadiabatic excitations for this model. 
Therefore, more precise adiabatic control is required to achieve accurate capture of ground-state correlation in experiments than in the first model. For models which in general does not possess the ground-state correlation robustness, our NQES algorithm also provides an avenue to assess the experimental performance of adiabatic ground-state preparation: Since we are able to compute the finite-field model, for each $h$ within the experimental regime, we can use the NQES algorithm to compute the energy gap between the ground state and the first excited state coupled to it by the transverse field ramp. Narrower gaps require slower ramps to maintain adiabaticity, which also provides guidance for the experimental protocol optimization.

Our algorithm offers a scalable and general-purpose framework for characterizing noisy intermediate-scale quantum devices. By providing access to both energy spectra and correlation functions of many-body quantum systems with long-range interactions, our method enables quantitative comparisons between theoretical predictions and experimental results. In particular, its ability to solve low-lying excited states proves valuable in diagnosing non-adiabatic errors in adiabatic protocols, as demonstrated in the trapped-ion setting. Also, it is appealing to exploit our algorithm for unsolved strongly interacting spin models, pinpointing their critical phenomena. The NQES algorithm thus functions not only as a numerical solver, but also as a versatile framework for validating and benchmarking programmable quantum simulators~\cite{Joshi2023Exploring,Iqbal2024NonAbelian}.

\vspace{.3cm}
\noindent\textbf{\color{sec_red}\large{}Discussions and outlooks}

From a computational complexity standpoint, the general spectral gap problem has been proven undecidable~\cite{Lloyd1993Quantummechanical,Cubitt2015Undecidability,Bausch2020Undecidability}, implying that no variational method\textemdash including ours\textemdash can be expected to resolve all spectral properties across arbitrary quantum systems. Nevertheless, within physically relevant regimes, our results demonstrate that the NQES algorithm can reliably capture multiple low-lying excited states in strongly interacting spin models at system sizes beyond the reach of conventional techniques. 
By incorporating physical prior knowledge through a tailored curriculum learning pipeline, we significantly enhance convergence and mitigate variational instabilities rooted in the sign problem.

The framework established here opens new directions at the interface of quantum many-body physics, artificial intelligence, and experimental benchmarking. First, the ability to compute multiple observables for several eigenstates within a unified training process offers a practical tool for validating quantum simulators and distinguishing physical signatures from preparation-induced artifacts. Second, the determinant-based formulation is agnostic to the choice of underlying network and could be extended to incorporate autoregressive, equivariant, tensorized architectures, and so on with improved expressivity~\cite{Carrasquilla2017Machine,vanNieuwenburg2017Learning}. 
Third, by engineering physics-informed curriculum learning that gradually introduces key parameters\textemdash interaction range, disorder strength, frustration angle, or symmetry constraints\textemdash this strategy could be extended to probe rich phenomena in other many-body systems, such as many-body localization in disordered spin chains, quantum spin-liquid phases in geometrically frustrated magnets, and topological order in lattice gauge theories. 

\vspace{.3cm}
\noindent\textbf{\color{sec_red}\large{}Methods}{\large\par}
\setcounter{figure}{0}
\renewcommand{\theHfigure}{A.Abb.\arabic{figure}}
\renewcommand{\figurename}{Extended Data Fig.}

\textbf{Efficient evaluation of the cost function}

To simultaneously access multiple low-lying eigenstates, we employ a variational framework that combines $K$ independent wavefunction ans\"atze $\{\psi_k\}_{k=1}^{K}$ into a single Slater-determinant function
\[
\Psi(\mathcal S)=\det[\mathbf{\Psi}(\mathcal S)], \text{where } \mathbf{\Psi}(\mathcal{S})_{k k'}=\psi_k(S^{k'}).
\]
Here, $\mathcal{S} = \{S^1, \dots, S^K\}$ denotes a set of spin configurations, each sampled according to the joint probability distribution $|\Psi(\mathcal{S})|^2$ during variational Monte Carlo updates.
Define $H\mathbf \Psi(\mathcal S)_{kk'}=\langle S^k|H|\psi_{k'}\rangle$. The cost function, corresponding to the total energy of the $K$ lowest eigenstates, can be expressed as the trace of a compact $K \times K$ local energy matrix:
\[
\langle \Psi|\mathcal H|\Psi\rangle/\langle \Psi|\Psi\rangle=\mathbb{E}_{\mathcal S \sim \Psi^2}\left[  \langle \mathcal S|\mathcal H|\Psi\rangle/\Psi(\mathcal S)\right]=\mathrm{Tr}[\mathbf E_{\mathrm{loc}}],
\]
where the local energy matrix 
\[
\mathbf E_{\mathrm{loc}}=\mathbb{E}_{\mathcal{S} \sim \Psi^2}\left[\boldsymbol{\Psi}(\mathcal{S})^{-1} H \boldsymbol{\Psi}(\mathcal{S})\right].\]

Parameter gradients are evaluated using standard restricted Boltzmann machine techniques and optimized via stochastic reconfiguration~\cite{Sorella2007Weak}.   The determinant structure ensures orthogonality among the $\psi_k$  throughout the optimization process.

At convergence, one obtains a local energy matrix $\mathbf E_{\mathrm{loc}}$ that approaches the diagonal matrix of the exact eigenvalues $\mathrm{diag}(E_0,\ldots,E_{K-1})$ up to a basis rotation.  Diagonalization of $\mathbf{E}_{\mathrm{loc}}$ thus yields the individual excitation energies. The same basis transformation can be applied to any other operator matrix $\mathbf{O}_{\mathrm{loc}}$, which is defined similarly as $\mathbf{E}_{\mathrm{loc}}$, to extract state-resolved expectation values $\langle \psi_i | O | \psi_i \rangle$ from diagonal terms.

\vspace{.2cm}
\textbf{Stochastic reconfiguration optimization}

We optimize the neural network by exploiting the stochastic reconfiguration method~\cite{Sorella2007Weak}. First, we introduce the scheme for learning the ground state. 
To enable gradient-based optimization, we define the variational derivative for a single wavefunction ans\"atz $\psi_W(S)$ with respect to the $l$-th network parameter $W_l$ as:
\begin{equation}
{D}_{l}(S)=\frac{\partial_{W_l} \psi_W(S)}{\psi_W(S)}.
\end{equation}
The local energy is given by:
\begin{equation}
 {E}_{\mathrm{loc}}({S})=\frac{\left\langle{S}|{H}| \psi_W\right\rangle}{\psi_W({S})}.
\end{equation}
During optimization, parameters are updated iteratively at step $p$ as:
\begin{equation}\label{eq:update}
W(p+1)=W(p)-\gamma(p) C^{-1}(p) F(p),
\end{equation}
with learning rate $\gamma(p)$, covariance matrix
\begin{equation} \label{eq:CovMatrix}
C_{ll'}(p) =  \overline{D_l^* D_{l'} } -\overline{D_{l}^{*}} \cdot  \overline {D_{l'}},
\end{equation}
and forces
\begin{equation} \label{eq:forces}
F_l(p) =\overline {E_{\mathrm{loc}} D_l^*}  - \overline {E_{\mathrm{loc}} }\cdot \overline {D_l^*} ,
\end{equation}
where $\overline  X=\mathbb{E}_{ S \sim \psi^2}[X(S)]$ denotes the classical average over quantity $X$ which can be evaluated via the Monte-Carlo sampling method. To ensure numerical stability when $C$ becomes ill-conditioned or singular, we regularize the matrix inversion by adding a small positive constant to its diagonal entries.

In our NQES algorithm, we generalize the above scheme to the task of optimizing $K$ neural networks simultaneously. In this scenario, the variational derivative with respect to the $l$-th parameter of the $k$-th network, denoted $\mathcal{W}_{k,l}$, is computed as:
\begin{equation}
  D_{k,l}(\mathcal{S})=
  \frac{\partial_{\mathcal{W}_{k,l}} \Psi\left(\mathcal{S}\right)}{\Psi\left(\mathcal{S}\right)}.
\end{equation}
These derivatives are concatenated into a vector of length $K$ times that of the single-state case for parameter update. The local energy for the expanded ans\"atz is taken as the trace over the local energy matrix:
\begin{equation}
  \mathcal{E}_\mathrm{loc}(\mathcal{S}) = \mathrm{Tr} \left[\boldsymbol{\Psi}^{-1}(\mathcal{S}) H \boldsymbol{\Psi}(\mathcal{S})\right].
\end{equation}

\vspace{.2cm}
\textbf{Local unitary transformation}

To improve convergence and enhance accuracy for the NQES procedure, we have adopted a curriculum learning~\cite{Bengio2009Curriculum} strategy. As a proxy for the long-range Haldane-Shastry model, we consider the antiferromagnetic Heisenberg chain ${H}^{\prime}$ with periodic boundary conditions, which captures key features of the target model. 
We initiate training on a transformed version of this model\textemdash the XXZ-type Hamiltonian by
\(
H_{\mathrm{XXZ}}= U^\dagger {H}^{\prime} U,
\) which is related to $H^{\prime}$ via a local unitary transformation $U = \prod_{i=1}^{N/2} \sigma_z^{2i}$. After obtaining a converged variational solution for $H_{\mathrm{XXZ}}$, we apply this transformation to the network to obtain the solution to $H^{\prime}$. 

For the restricted Boltzmann machine representation, this unitary operation induces a simple analytical reparameterization: parameters $a_{2i}$ on even sites are shifted by $-\frac{\mathrm{i}\pi}{2}$,
\[
 a'_{2i} = a_{2i} - \frac{{\rm i} \pi}{2},i=1,\cdots,\frac{N}{2},
\]
while all other parameters remain unchanged. This procedure allows us to transfer the learned structure from the XXZ model to the antiferromagnetic Heisenberg model without retraining from scratch, enabling us to subsequently continue training toward the Haldane-Shastry model.

\vspace{.2cm}
\textbf{Memory-efficient Krylov subspace method}

To overcome the memory bottlenecks inherent in stochastic reconfiguration, we adopt a matrix-free solver based on Krylov subspace techniques. Specifically, storing the full covariance matrix $C$ becomes infeasible for large systems: its memory requirement scales quadratically with the number of variational parameters and reaches the terabyte range for systems with 32 spins on computing setups with 64 CPU cores.

To circumvent this limitation, we apply the MINRES-QLP algorithm~\cite{Choi2011MINRESQLP} to solve the linear system involving the covariance matrix $C$ without explicitly constructing or storing it. Instead, it only requires a routine that computes the matrix-vector product $C \cdot v$, making it ideally suited for matrix-free implementations. 
During each Krylov iteration, we directly evaluate this matrix-vector product from Monte Carlo samples of the derivative observables $D_l(C)$, thereby eliminating the need to store or manipulate the full matrix. Although this approach incurs additional floating-point computations, it substantially reduces main-memory traffic. As a result, for systems with $N \geq 16$, the matrix-free implementation becomes more efficient than traditional methods on modern multi-core architectures. By eliminating the dominant memory bottleneck, this method enables the stochastic reconfiguration optimization to scale to significantly larger quantum systems, all within a fixed hardware budget.

\vspace{.2cm}
\textbf{Compact encoding scheme for the spin configurations}

To further accelerate the Monte Carlo evaluation of quantum observables, we introduce a compact binary encoding scheme for spin configurations and the associated Pauli operator actions. Each spin state in the computational basis is encoded as a bit string, with $\ket{\uparrow} \rightarrow 0$ and $\ket{\downarrow} \rightarrow 1$. For example, the state $\ket{\uparrow\downarrow\downarrow\uparrow\uparrow}$ is represented by the binary string $01100$.

Under this encoding, the action of Pauli operators can be implemented efficiently using bitwise operations. Specifically, the operator $\sigma_z^i$ multiplies the state by $(-1)^{b_i}$, where $b_i$ denotes the $i$-th bit; $\sigma_x^i$ flips the $i$-th bit; and $\sigma_y^i$ acts as a composite operation owing to the identity $\sigma_y = i \sigma_x \sigma_z$. This encoding significantly streamlines the evaluation of Hamiltonian matrix elements. For instance, for the periodic transverse-field Ising Hamiltonian $H \sim \sum_i \sigma_z^i \sigma_z^{i+1}$, diagonal matrix elements can be computed using only three low-level operations: a left bit shift, an exclusive-or, and a population count which counts the number of 1s in a binary word:
\begin{align*}
& \bra{\uparrow\downarrow\downarrow\uparrow\uparrow} H \ket{\uparrow\downarrow\downarrow\uparrow\uparrow} = \bra{01100} H \ket{01100} \\
&= N - 2\cdot \texttt{POPCOUNT}(\texttt{XOR}(01100, \texttt{SHL}(01100, 1)))
\end{align*}
where $N=5$. Here $\texttt{SHL}(01100, 1)$ left-shifts $01100$ by one bit:
\[
\texttt{SHL}(01100, 1) = 11000\,,
\]
$\texttt{XOR}(a,b)$ computes the exclusive-or of the two bit strings:
\[
\texttt{XOR}(01100, 11000) = 10100\,,
\]
and $\texttt{POPCOUNT}(a)$ counts the number of ones in the bit string:
\[
\texttt{POPCOUNT}(10100) = 2.
\]
These bit-level operations are directly translated to single-instruction, multiple-data instructions by modern optimizing compilers, allowing multiple spin updates to be computed in parallel within a single instruction. This scheme yields a substantial computational speedup, especially in Monte Carlo sampling where Hamiltonian evaluations are required at each step. For further performance metrics and hardware specifications, the information about the utilized computational resources is exhibited in Supplementary Section~V.

\vspace{.3cm}
\noindent\textbf{\color{sec_red}\large{}Data availability}

The data presented in the figures and that support the other findings of this study will be publicly available at Zenodo.org upon publication. 

\vspace{.3cm}
\noindent\textbf{\color{sec_red}\large{}Code availability}

All the relevant source codes will be publicly available at Zenodo.org upon publication. 

\vspace{.3cm}
\noindent\textbf{\color{sec_red}\large{}Acknowledgements}

We thank Z. Lu, D. Luo, D. Yuan, Q. Ye, and S. Jiang for discussions. This work is supported by the National Natural Science Foundation of China (grant nos.~T2225008 and 12075128), the Innovation Program for Quantum Science and Technology (grant nos.~2021ZD0301601, 2021ZD0301605, and 2021ZD0302203), the Shanghai Qi Zhi Institute Innovation Program (grant nos.~SQZ202318 and SQZ202317), the Tsinghua University Dushi Program, the Tsinghua University Initiative Scientific Research Program, and the Ministry of Education of China. L.-M.D. acknowledges in addition support from the New Cornerstone Science Foundation through the New Cornerstone Investigator Program.

\vspace{.3cm}
\noindent\textbf{\color{sec_red}\large{}Author contributions}

Y.W. and D.-L.D. proposed and supervised the project with support from L.-M.D.. Y.M. and C.L. developed the theory and conducted numerical simulations with support from W.L. and S.-Y.Z.. All authors contributed to the analysis of data, discussions of the results, and writing of the manuscript.

\vspace{.3cm}
\noindent\textbf{\color{sec_red}\large{}Competing interests}  

Y.W. and L.-M.D. hold shares with HYQ Co. The other authors declare no competing interests.

\bibliography{QAI}

\end{document}


\title{Supplementary: Solving excited states for long-range interacting trapped ions with neural networks\\
}

\captionsetup[figure]{name={Supplementary Figure}, singlelinecheck=off,labelfont=bf,labelsep=period,justification=raggedright} 
\captionsetup[table]{name={Supplementary Table}, singlelinecheck=off,labelfont=bf,labelsep=period,justification=raggedright}

\renewcommand{\bibnumfmt}[1]{[S#1]}
\renewcommand{\citenumfont}[1]{S#1}
\setcounter{equation}{0}
\setcounter{figure}{0}
\setcounter{table}{0}

\renewcommand{\theequation}{S\arabic{equation}}
\renewcommand{\thefigure}{S\arabic{figure}}

\setcounter{figure}{0}
\setcounter{table}{0}
\renewcommand\thefigure{S\arabic{figure}}
\renewcommand\thetable{S\arabic{table}}

\maketitle
\tableofcontents

\section{Representing quantum states with restricted Boltzmann machines}

The wavefunction of many-body spin systems, denoted as $\psi$, can be thought of as a mapping from a spin configuration ${S}=\{s_1, s_2, ..., s_N\},s_i=\pm1,i=1,2,\cdots N$ to a complex number. As the system size $N$ increases, the number of possible configurations increases exponentially, and so does the number of parameters needed to fully encode the information of a generic quantum state. As a result, representing a many-body state on a classical computer is generally difficult.

One approach to mitigate this problem is to use neural networks, which can approximate the mapping with significantly fewer parameters. One such neural network architecture of interest to us is the restricted Boltzmann machine (RBM). The RBM consists of a visible layer (physical nodes) and a hidden layer (unphysical nodes), with inter-layer connections between them. The network is restricted in the sense that connections between visible-visible and hidden-hidden nodes are not allowed. This restriction facilitates a practical algorithm for learning the network parameters. 

To compute the wavefunction for inputs from the visible layer, the RBM employs the following equation~\cite{Carleo2017Solving}: 
\begin{equation}
    \psi_W(S)=\sum_{\{h_j\}}e^{\sum_i a_i s_i+\sum_{j} b_j h_j+\sum_{i j} w_{i j} s_i h_j}, 
    \label{RBM}
\end{equation}
where $h_j\in\{-1,1\}$, and the network parameters ${W}=\{a_i, b_j,w_{ij}\}$. In general, these parameters are complex numbers. Tracing out the hidden variables, the wavefunction can be computed on the computational basis as \begin{equation}
\psi_{W}({S})=e^{\sum_i a_i s_i} \times \prod_{j} \theta_j(S),
\end{equation}
where $
\theta_j({S})=2 \cosh \left[b_j+\sum_i w_{{i}{j}} s_i\right]$. Then we can represent the wavefunction of a physical state as 
\begin{equation}
   \left|\psi_W\right\rangle= \sum_{S}\psi_{W}({S})\left|S\right\rangle,
\end{equation}
 where the sum is over all possible spin configurations.
\section{Learning energy eigenstates}
Given the system Hamiltonian, we discuss efficient optimization algorithms for energy eigenstates. In Section~\ref{GS}, we introduce the method for learning the ground state. After that, we extend the method to find excited states in Section~\ref{ES}. 

\subsection{Initial weight distributions}
The initial RBM weights are generated randomly from a Gaussian distribution of zero mean and the following variances
\[
\sigma^2(a) = \frac{1}{N}\qquad\textrm{for $a$},
\]
\[
\sigma^2(b) = \frac{1}{M}\qquad\textrm{for $b$},
\]
and
\[
\sigma^2(w) = \frac{1}{NM}\qquad\textrm{for $w$}.
\]
We make this choice so that the wavefunction built from these initial weights is of order unity on average, mitigating the risk of the wavefunction diverging during optimization for large system sizes. 
\subsection{Learning the ground state}\label{GS}
When training a neural network, we need to predefine a cost function. The cost function is a function of the network parameters, which measures the distance between the current output and the target output.
During the optimization process, it is minimized to obtain the desired network parameters. To find the ground state, we set the cost function to be the energy expectation value $E({W})=\left\langle\psi|{H}| \psi\right\rangle /\left\langle\psi\mid\psi\right\rangle$, followed by a variational method to minimize $E({W})$ so that the representation of the ground state is obtained.

\subsubsection{Optimization scheme for the ground state}\label{GSoptm}
To optimize the variational wavefunction, we employ the stochastic reconfiguration (SR) method, which can be understood as a variational approximation to an imaginary-time evolution. In this framework, the quantum state $\psi_W$ is evolved under an imaginary time $\mathrm{i}\tau$ to approximate $e^{-\tau H} \psi_W$, a process that drives the state toward the ground state as $\tau \rightarrow \infty$.

Since $\psi_W$ is represented by a neural network with parameters $W$, this continuous evolution must be mapped onto a discrete update of the parameters. During each iteration step, we shift the network parameters $W$ by $\delta W$ such that the updated state $\psi_{W + \delta W}$ best approximates $e^{-\delta \tau H} \psi_W$ within the variational manifold, where $\delta \tau$ is a small time interval. A detailed derivation of this procedure is provided in ref.~\cite{Nomura2024Boltzmann}. The resulting update rule is presented as follows.  First, we introduce the variational derivatives with respect to the $k$-th network parameter
\begin{equation}
{D}_{{k}}({S})=\frac{1}{\psi_W({S})} \partial_{{W}_{{k}}} \psi_W({S}),\end{equation} and the local energy \begin{equation}
    {E}_{\mathrm{loc}}({S})=\frac{\left\langle{S}|{H}| \psi_W\right\rangle}{\psi_W({S})}.
\end{equation}The SR updates the parameters at the $p$-th iteration as follows:
\begin{equation}
{W}(p+1)={W}(p)-\gamma(p) C^{-1}(p) F(p),
\end{equation}
with learning rate $\gamma(p)$, the covariance matrix
\begin{equation} \label{eq:CovMatrix}
C_{ll^{\prime}}(p) =  \overline{D_l^* D_{l^{\prime}} } -\overline{D_{l}^{*}} \cdot  \overline {D_{l^{\prime}}},
\end{equation}
and forces
\begin{equation} \label{eq:forces}
F_l(p) =\overline {E_{\mathrm{loc}} D_l^*}  - \overline {E_{\mathrm{loc}} }\cdot \overline {D_l^*} ,
\end{equation}
where $\overline  X=\mathbb{E}_{ S \sim \psi^2}[X(S)]$ denotes the classical average over quantity $X$. Since $C$ may be non-invertible, we apply an explicit regularization, adding a small number to the diagonal term of the matrix. 
\subsubsection{Efficient sampling of configurations}
During the training process, the average of quantities $\overline  X=\mathbb{E}_{ S \sim \psi^2}[X(S)]$ can be computed through the standard Monte Carlo paradigm. First, we apply the Metropolis-Hastings method to generate a Markov chain of configurations ${S}_1,...,{S}_P$ that obeys the probability distribution $|\psi_W({S})|^2$.  Specifically, starting from a configuration ${S}_n$, the configuration ${S}_{n+1}$ with one of the spins flipped is accepted according to the probability
\begin{equation}
A\left({S}_n\rightarrow {S}_{n+1}\right)=\min \left(1,\left|\frac{\psi_W\left({S}_{n+1}\right)}{\psi_W\left({S}_{n}\right)}\right|^2\right).
\end{equation}

Next, we can determine the values in equation~\eqref{eq:CovMatrix} and equation~\eqref{eq:forces} by calculating the variational derivatives and local energy for each configuration within the Markov chain and then taking the average of these quantities. 

As the system size increases, the statistical requirements for Monte Carlo sampling grow significantly, necessitating a proportional increase in the number of samples to maintain accuracy. To mitigate the computational overhead, we parallelize the sampling task across multiple CPU cores. Each core independently generates a Markov chain, thereby distributing the workload and accelerating data generation. To ensure statistical independence from initialization biases, each chain must undergo a number of thermalization steps before collecting samples. 

Note that when one spin is flipped during Monte Carlo sampling, we do not need to recompute the wavefunction. Instead, it can be updated as follows:
\begin{equation}
  \psi_W({S}_{n+1}) = \psi_W({S}_{n}) e^{-2 a_i s_i}\prod_{j=1}^M \left[\cosh(2w_{ij} s_i) - \tanh\theta_j \sinh(2w_{ij}s_i)\right].
\end{equation}
Here the $i$th spin is flipped from ${S}_n$ to ${S}_{n+1}$ . This requires storing the $\tanh\theta_j$ vector, which we simultaneously update to
\begin{equation}
   \tanh\theta_j \leftarrow \frac{\tanh\theta_j \cosh(2w_{ij}s_i) - \sinh(2 w_{ij}s_i)}
{\cosh(2w_{ij}s_i) - \tanh\theta_j\sinh(2w_{ij}s_i)}.
\end{equation}
The variational derivatives can be efficiently evaluated according to the RBM ans\"atz in equation~\eqref{RBM} as follows:
\begin{equation}
\begin{aligned}
& \frac{1}{\psi_W({S})} \partial_{a_i} \psi_W({S})= s_i, \\
& \frac{1}{\psi_W({S})} \partial_{b_j} \psi_W({S})=\tanh \left[\theta_j({S})\right], \\
& \frac{1}{\psi_W({S})} \partial_{w_{i j}} \psi_W({S})=s_i \tanh \left[\theta_j({S})\right].
\end{aligned}
\end{equation}

\subsection{Learning excited states}\label{ES}
\subsubsection{Optimization scheme for excited states}
In this section, we present the scheme to efficiently evaluate the cost function for multiple excited states. 
We define the matrix for the composite function
\begin{equation}
\mathbf{\Psi}(\mathcal{S}) = \left[\begin{array}{ccc}
\psi_1({S}^1) & \cdots & \psi_K\left({S}^1\right) \\
\vdots & & \vdots \\
\psi_1\left({S}^K\right) & \cdots & \psi_K\left({S}^K\right)
\end{array}\right].
\end{equation}
This determinant is equal to the composite function
\begin{equation}
\Psi(\mathcal{S}) =\mathrm{det}(\mathbf{\Psi}\left(\mathcal{S}\right)),
\label{overallwavef}
\end{equation}
where $\mathcal{{S}}=\{{S}^1, \ldots, {S}^K\}$ are $K$ different spin configurations.  $\psi_1, ...,\psi_K$ are $K$ single-state RBM ans\"atze for ${H}$ with network parameters $\mathcal{W}=\{{W}_1,{W}_2, ..., {W}_K\}$. 

For any physical quantity $O$, we have the following equation~\cite{Pfau2024Accurate}
\begin{equation}
    \mathbb{E}_{\mathcal S \sim\Psi^2}\left[\mathbf{\Psi}(\mathcal{S})^{-1} {{O}} \boldsymbol{\Psi}(\mathcal{S})\right]=\mathbf{S}^{-1} {\mathbf{O}},
\end{equation}
 where
 \begin{equation}
     {O}\mathbf{\Psi}\left(\mathcal{S}\right) =\left[\begin{array}{ccc}
\left\langle {S}^1|{O}| \psi_1\right\rangle & \cdots & \left\langle  {S}^1|{O}| \psi_K\right\rangle\\
\vdots & & \vdots \\
\left\langle  {S}^K|{O}| \psi_1\right\rangle & \cdots & \left\langle  {S}^K|{O}| \psi_K\right\rangle
\end{array}\right],
 \end{equation}
 \begin{equation}
     \mathbf{S} =\left(\begin{array}{ccc}
\left\langle\psi_1|\psi_1\right\rangle & \ldots & \left\langle\psi_1 |\psi_K\right\rangle \\
\vdots & & \vdots \\
\left\langle\psi_K |\psi_1\right\rangle & \ldots & \left\langle\psi_K|\psi_K\right\rangle
\end{array}\right),
 \end{equation}
 and
 \begin{equation}
     {\mathbf{O}} =\left(\begin{array}{ccc}
\left\langle\psi_1|{O}| \psi_1\right\rangle & \ldots & \left\langle\psi_1|{O}| \psi_K\right\rangle \\
\vdots & & \vdots \\
\left\langle\psi_K |{O}|  \psi_1\right\rangle & \ldots & \left\langle\psi_K |{O}|  \psi_K\right\rangle
\end{array}\right).
 \end{equation}
As discussed in the main text, our optimization goal is to minimize the total energy $\mathbb{E}_{\mathcal S \sim \Psi^2}\left[  \langle \mathcal S|\mathcal H|\Psi\rangle/\Psi(\mathcal S)\right]$. It can be proven that it equals $\mathrm{Tr}[\mathbf{S}^{-1} {\mathbf{H}}]$. Thus, we can generalize the local energy from a scalar to a matrix:
\begin{equation}
\mathbf{E}_\mathrm{loc} = \mathbb{E}_{\mathcal{S}\sim \Psi^2}\left[\boldsymbol{\Psi}(\mathcal{S})^{-1} {H} \boldsymbol{\Psi}(\mathcal{S})\right],
\end{equation} 
and estimate its trace $\mathrm{Tr}[\mathbf{E_{\mathrm{loc}}}]$ as the cost function. Since the total ans\"atz is a Slater determinant, $\psi_1, ...,\psi_K$ will not collapse onto each other. 

If $\psi_1,\cdots,\psi_K$ are the $K$ lowest energy eigenstates $\psi_1^*,\cdots,\psi_K^*$, we have 
\begin{equation}
    \boldsymbol{\Psi}^{  \boldsymbol{*}-1 } {H} \boldsymbol{\Psi^*}= \boldsymbol{\Psi}^{  \boldsymbol{*}-1 }  \boldsymbol{\Psi^*}\Lambda=\Lambda,
\end{equation}
where $\Lambda=\mathrm{diag}(E_0,\cdots,E_{K-1})$ presents the $K$ lowest eigenenergies. The composite function $\Psi^*$ is without doubt a ground state of the expanded Hamiltonian $\mathcal{H}$ with the eigenvalue of $\sum_iE_i$ on this ans\"atz. However, it is not the only ground state. If we linearly transform the matrix $\mathbf{\Psi}^*$ to $\mathbf{\Psi}^*A$ , we have
\begin{equation}
    (\boldsymbol{\Psi^*}A)^{-1} {H} \boldsymbol{\Psi^*}A=  A^{-1}  \Lambda A,
\end{equation}
from which we get the same energy $\mathrm{Tr}[\mathbf{E_{\mathrm{loc}}}]=\sum_iE_i$ and thus it is also a global minimum to the optimization problem. This is a more general solution which we typically obtain. In this case, we can diagonalize the matrix $\mathbf{E}_\mathrm{loc}=U^{-1}\Lambda U$ to obtain eigenenergies for the single system from its diagonal terms. The transformation is $U=A^{-1}\Sigma$, where $\Sigma$ is an undetermined diagonal matrix. 

Expectations of other physical quantities $O$ for excited states can be calculated as well.  First, we define the local matrix for $O$ as $\mathbf{O}_\mathrm{loc}=\mathbb{E}_{\mathcal S \sim\Psi^2}\left[\mathbf{\Psi}^{-1}{{O}} \boldsymbol{\Psi}\right]$. When $\mathbf{\Psi}=\mathbf{\Psi}^*A$, we have $ \mathbf{O}_\mathrm{loc}=\mathbf{S}^{-1} {\mathbf{O}}=\mathbf{A}^{-1} {\mathbf{O}}^{\star} \mathbf{A}$. By transforming $\mathbf{O}_\mathrm{loc}$ with $U$ which we obtained earlier to diagonalize $\mathbf{E}_\mathrm{loc}$, we get $\boldsymbol{\Sigma}^{-1} \hat{\mathbf{O}}^{\star} \boldsymbol{\Sigma}$. Its diagonal terms correspond to $\left\langle\psi_i^* |{O}|  \psi_i^*\right\rangle , i=1,2,\cdots,K$. In the main text, we calculate the excited-state correlations $\left\langle\sigma_z^i\sigma_z^j\right\rangle ,i,j=1,\cdots,N$ in this way.

\subsubsection{Efficient sampling of configurations}
In comparison to the single-state scenario, the number of network parameters has increased by a factor of $K$. Additionally, there is a shift in the cost function to $\mathrm{Tr}[\mathbf{E_{\mathrm{loc}}}]$. These changes are reflected in the SR method when updating the network parameters. 

We also apply the Metropolis-Hastings method to generate a Markov chain of configurations $\mathcal{S}_1,...,\mathcal{S}_P$ that obeys the probability distribution $|\Psi_\mathcal W(\mathcal{S})|^2$.  Starting from a configuration $\mathcal{S}_n$, the configuration $\mathcal{S}_{n+1}$ with one of the spins flipped  is accepted according to the probability
\begin{equation}
A\left(\mathcal{S}_n\rightarrow \mathcal{S}_{n+1}\right)=\min \left(1,\left|\frac{\Psi\left(\mathcal{S}_{n+1}\right)}{\Psi\left(\mathcal{S}_{n}\right)}\right|^2\right).
\end{equation}
Next, we can determine the values of equation~\eqref{eq:CovMatrix} and equation~\eqref{eq:forces} by averaging the variational derivatives $D_{k,l}$ and local energy $\mathrm{Tr}[\mathbf{E_{\mathrm{loc}}}]$ for the Markov chain samples.

The variational derivative with respect to the $k$-th parameter of the $l$-th RBM $\psi_l$ is evaluated as
\begin{equation}
D_{k,l}(\mathcal{S})=\frac{1}{\Psi\left(\mathcal{S}\right)} \partial_{\mathcal{W}_{k,l}} \Psi\left(\mathcal{S}\right).\end{equation} 
Expanding the Slater determinant
\begin{equation}
    \Psi(\mathcal{S})=\sum_{j_1  ... j_K}(-1)^{N(j_1 ... j_K)}\psi_1({S}^{j_1})...\psi_K({S}^{j_K}),
\end{equation}
we get 
\begin{equation}
\begin{aligned}
D_{k,l}(\mathcal{S}) & =\Psi\left(\mathcal{S}\right)^{-1}\cdot\partial_{\mathcal{W}_{k,l}}\sum_{j_1 ... j_K}(-1)^{N(j_1 ... j_K)}\psi_1({S}^{j_1})...\psi_K({S}^{j_K}) \\
& =\Psi\left(\mathcal{S}\right)^{-1}\cdot\sum_{j_1 ... j_K}(-1)^{N(j_1 ... j_K)}\psi_1({S}^{j_1})...\partial_{\mathcal{W}_{k,l}} \psi_l({S}^{j_l})...\psi_K({S}^{j_K})\\
&=\Psi\left(\mathcal{S}\right)^{-1}\cdot\mathrm{det}\left[\begin{array}{ccccc}
\psi_1({S}^1) & \cdots & \partial_{\mathcal{W}_{k,l}} \psi_l({S}^1)&\cdots&\psi_K({S}^1) \\
\vdots & &\vdots&&\vdots \\
\psi_1({S}^K) & \cdots &\partial_{\mathcal{W}_{k,l}} \psi_l({S}^K)&\cdots &\psi_K({S}^K)
\end{array}\right],
\end{aligned}
\end{equation}
where $N(j_1 ... j_K)$ is the inversion number of the sequence $j_1 ... j_K$. The variational derivatives for the single-state ans\"atz $\partial_{\mathcal{W}_{k,l}} \psi_l\left({S}\right)$ can be evaluated efficiently as in Section~\ref{GSoptm}. The single-state derivatives are concatenated into a vector $K$ times longer than in the single-state case for parameter update.

\section{Curriculum learning pipeline for a long-range antiferromagnetic model}
\begin{figure}[t]
    \centering
    \includegraphics[width=0.5\linewidth]{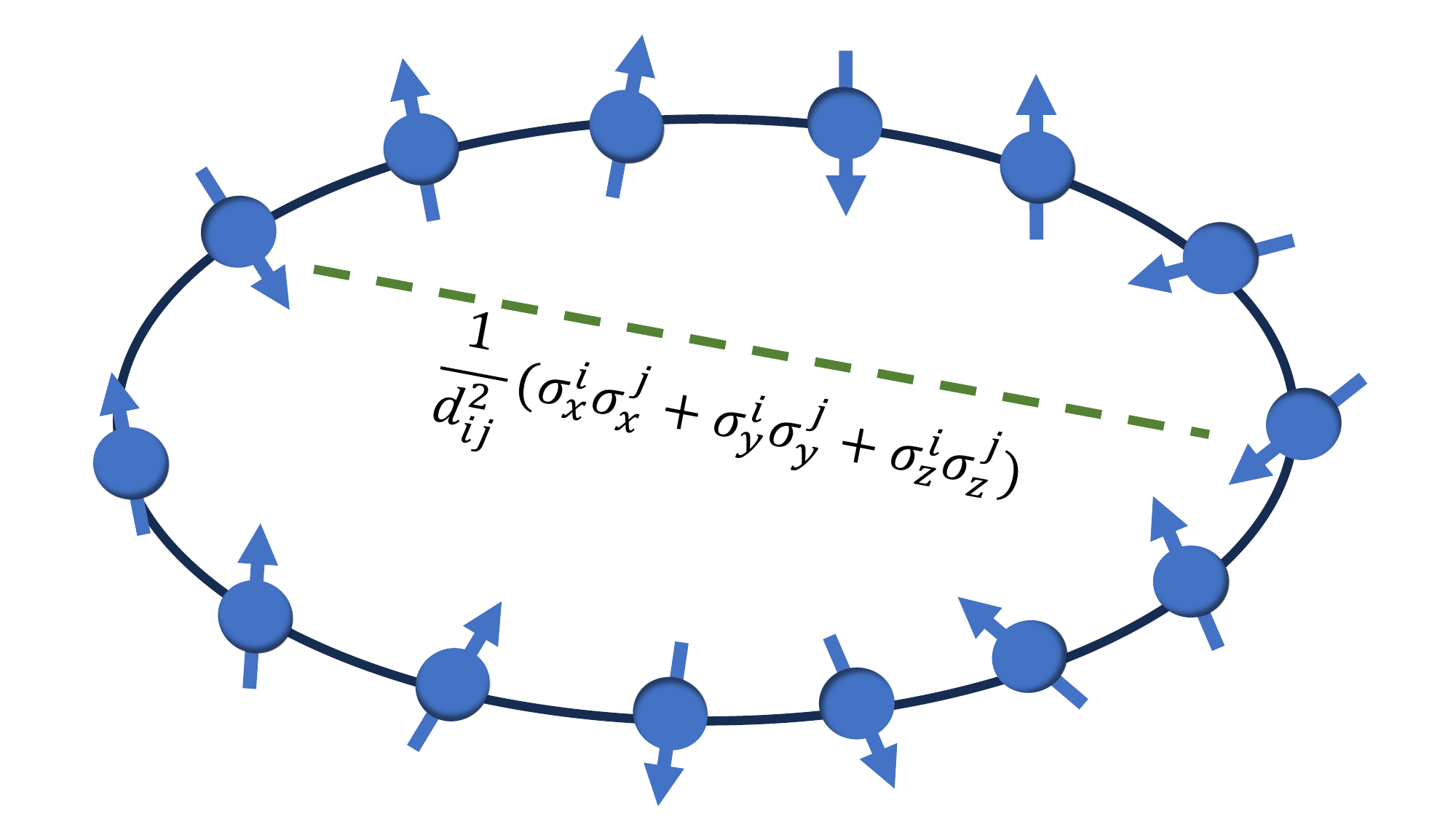}\caption{Illustration of the Haldane-Shastry model. Spins are arranged in a circle. Each spin is antiferromagnetically coupled to all other spins. The coupling strength decays according to a power law with distance.}
    \label{figS1}
\end{figure}
We benchmark our algorithm on the Haldane-Shastry model 
\begin{equation}
{H}_{1}=\sum_{i<j} \frac{1}{d_{ij}^2}\left({\sigma}^i_x {\sigma}^j_x+{\sigma}^i_y {\sigma}^j_y+{\sigma}^i_z{\sigma}^j_z\right),
\end{equation}
which is a spin-chain model with long-range antiferromagnetic interactions, as shown in Fig.~\ref{figS1}.

\subsection{Theoretical solution}
The Haldane-Shastry model is exactly solvable. In particular, the energy spectrum and eigenstates can be computed analytically. The eigenenergies of the Haldane-Shastry model are~\cite{Shastry1988Exact,Haldane1988Exact}:
\begin{equation}
\lambda _M/N=(\frac{1}{6}-\frac{1}{2}\mu +\frac{1}{6}\mu^3-\frac{1+4\mu}{6N^2})\pi^2,
\end{equation}
where $M $ is a non-negative integer no greater than $N/2$, and $\mu=2M/N\in [0,1]$. The ground state energy is obtained for $M=N/2$, and the first excited state energy is obtained for $M=N/2-1$.

\subsection{Sign problem of the Haldane-Shastry model}

A central task in quantum Monte Carlo is to estimate the quantum-state average of a physical observable $O$:
\begin{equation}
    \left\langle O\right\rangle=\frac{\sum_x |\psi(x)|^2\frac{1}{\psi(x)}\sum_{x'}O_{x x'}\psi(x')}{\sum_x |\psi(x)|^2}.
\end{equation}
It can be converted to a classical Monte Carlo problem $ \mathbb{E}_{x \sim\psi^2}[O_{\mathrm{loc}}]$, with $O_{\mathrm{loc}}=\frac{1}{\psi(x)}\sum_{x'}O_{x x'}\psi(x')$.

For the eigenenergy problem, the physically relevant $O$ is the Hamiltonian $H$. If all $H_{x x'}\le0,x\ne x'$ (stoquastic), we can set $\psi(x)>0$ for all $x$, and the variance for estimating $\left\langle H\right\rangle$ is stable. Otherwise, the positive and negative contributions in the sum over $E_{\mathrm{loc}}$ cancel out,  leading to an increase in the variance. In this case, the Hamiltonian suffers from the so-called sign problem. 

The sign problem is basis-dependent\textemdash it refers to the quantity of the Hamiltonian on a chosen basis. Therefore, if we can find an orthogonal basis on which the Hamiltonian is stoquastic, the sign problem is removed. And if we move back to the original basis, $H_{x x'}$ and $\frac{\psi(x')}{\psi(x)}$ would have the opposite sign. In other words, the sign structure of the Hamiltonian matrix is imprinted in the wavefunction. 

On the computational basis, it is easy to see that the Hamiltonian matrix of the Haldane-Shastry model is non-stoquastic: positive non-diagonal terms exist, which are determined by the interaction form ${\sigma}_x^i {\sigma}_x^j+{\sigma}_y^i {\sigma}_y^j+{\sigma}_z^i {\sigma}_z^j$. Therefore, the Haldane-Shastry model suffers from the sign problem in this basis. Previous works have proved that there does not exist single-qubit unitary operators $U_1 \otimes U_2 \otimes \cdots \otimes U_N$ to transform the Hamiltonian matrix into a stoquastic one~\cite{Bravyi2023Rapidly}.

\subsection{Curriculum learning scheme}

\begin{figure}[t]
    \centering
    \includegraphics[width=0.9\linewidth]{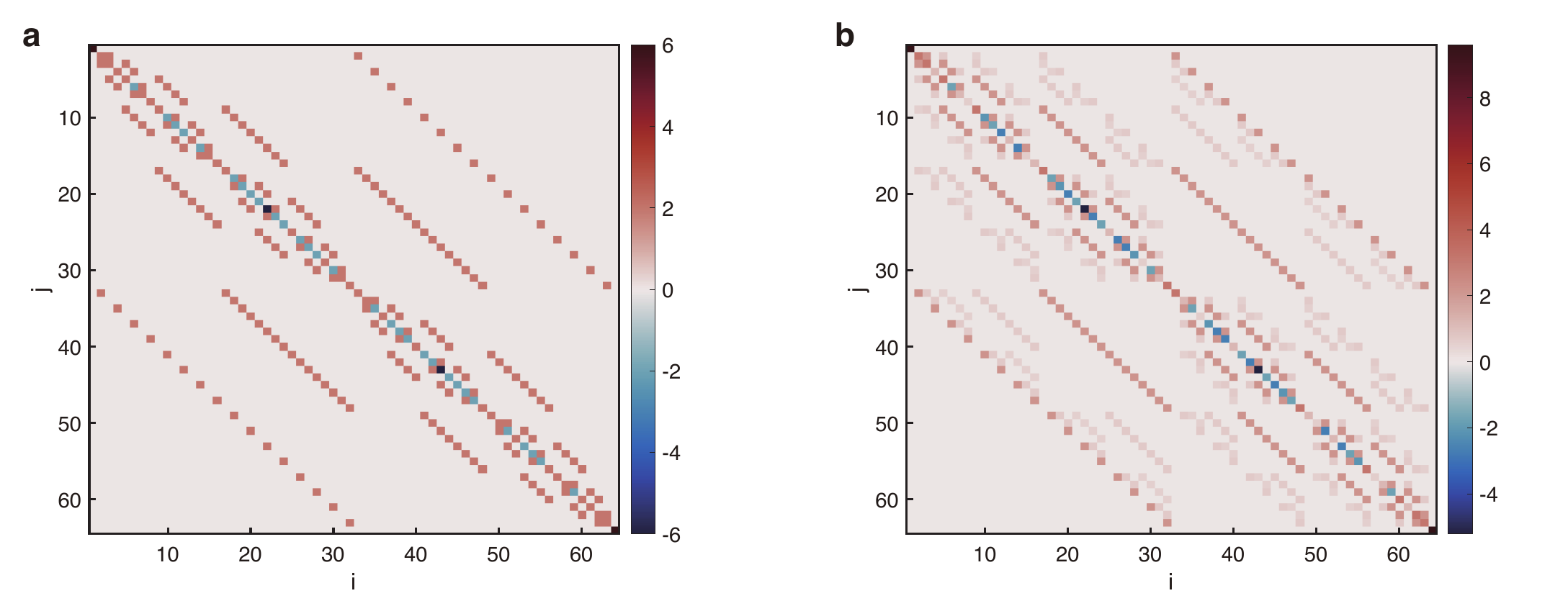}\caption{The Hamiltonian matrix on the computational basis of \textbf{a}, the antiferromagnetic Heisenberg model and \textbf{b}, the Haldane-Shastry model for 6 spins. Both models suffer from severe sign problems on this basis. $i$ and $j$ span the entire Hilbert space of dimension 64. }
    \label{figS2}
\end{figure}
The local-interaction counterpart of the Haldane-Shastry model, the antiferromagnetic Heisenberg model with periodic boundary conditions
\begin{equation}
H^{\prime}=\sum_{i}{\sigma}_x^i {\sigma}_x^{i+1}+{\sigma}_y^i {\sigma}_y^{i+1}+{\sigma}_z^i {\sigma}_z^{i+1},
\end{equation}
suffers from the same sign problem. In Fig.~\ref{figS2}\textbf{a, b}, we plot their Hamiltonian matrices, respectively. The largest off-diagonal elements of $H_1$ appear at the same matrix positions as the off-diagonal elements of  $H^{\prime}$. What differs, however, is that the antiferromagnetic Heisenberg model is known to be sign-problem-free\textemdash there exists a basis to make $H^{\prime}$ stoquastic. Hence, if we first obtain the antiferromagnetic Heisenberg model's wavefunction on the computational basis\textemdash which faithfully captures the sign pattern of its Hamiltonian as claimed earlier\textemdash the resulting phase information can serve as an initial ans\"atz for the Haldane-Shastry model, thereby alleviating the sign problem by ensuring that the dominant summands are negative.

Along this line, we resort to curriculum learning to mitigate the sign problem of the Haldane-Shastry model. First, we use a unitary operator $U$ to transform the antiferromagnetic Heisenberg model into the XXZ model
\begin{equation}
    H_{\mathrm{XXZ}}= U^\dagger H^{\prime} U = \sum_{i} -{\sigma}_x^i {\sigma}_x^{i+1}-{\sigma}_y^i {\sigma}_y^{i+1}+{\sigma}_z^i {\sigma}_z^{i+1},
\end{equation}
where $U = \prod_{i=1}^{N/2} \sigma^{2i}_z$.

The XXZ model is stoquastic in the computational basis, thus it is much easier to obtain its solutions. Once we have obtained the eigenstate $\psi$ to the XXZ model, we transform it to $U\psi$, which is the solution to the antiferromagnetic Heisenberg model. We are using the RBMs as variational ans\"atze. The action of $U$ on the RBMs can be easily transformed into an action on the network parameters $W$: 
\begin{equation}
\psi'=U\psi\quad\Rightarrow\quad a'_{2j} = a_{2j} - \frac{{\rm i} \pi}{2},j=1,\cdots,N/2.
\end{equation}
Starting from $\psi'$, we continue training with the full Haldane-Shastry Hamiltonian. Since the network already captures the dominant phase structure, learning now focuses on activating the additional long-range links unique to the Haldane-Shastry model, thereby markedly reducing the severity of the sign problem.

\section{Solving the many-body Hamiltonian of strongly interacting trapped ions}
We summarize experimental settings and parameters in the ref.~\cite{Guo2024Siteresolved} in this section. We will briefly introduce how the ions are trapped and how they are located in space. Then we discuss how a pair of global-manipulating lasers induces the Ising coupling between ions.  The method of generating the Hamiltonian data for our NQES algorithm is also presented.
\subsection{Trapping the 2D ion crystal}

A monolithic 3D Paul trap~\cite{Brownnutt2006Monolithic,Wang2020Coherently} with an RF electrode is used to hold 2D crystals of ${}^{171}\mathrm{Yb}^+$ ions. The voltage on the $4\times7=28$ DC segments can be controlled independently, together with an overall DC bias on the RF electrodes. This setting provides us with sufficient degrees of freedom to control the shape of the 2D crystal and to eliminate its micromotion perpendicular to the plane. The ion crystal is shaped into roughly an ellipse by adjusting the voltages on the electrodes.

\begin{figure}[t]
    \centering
    \includegraphics[width=0.9\linewidth]{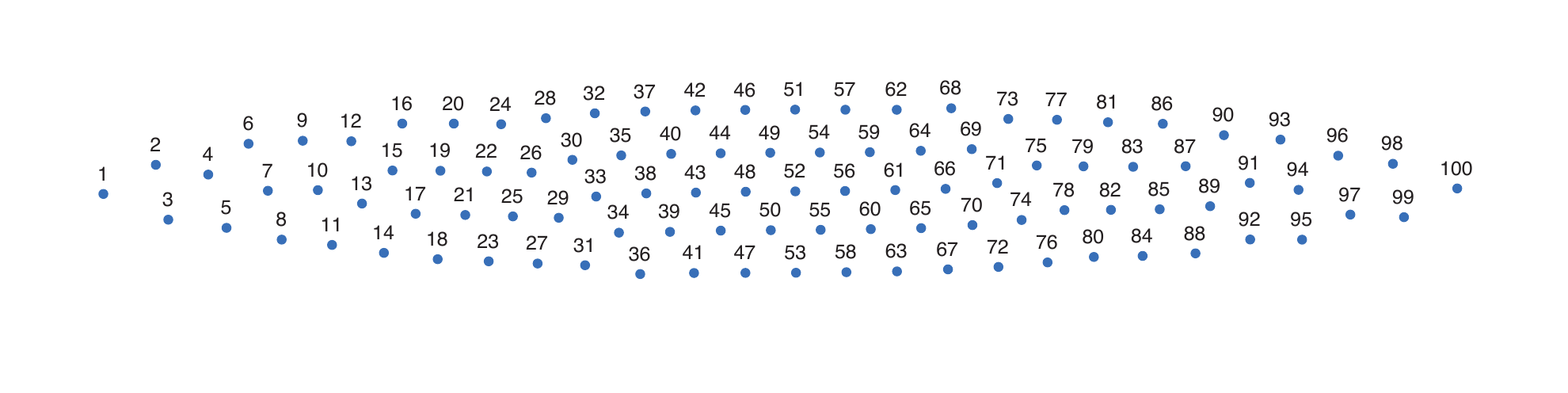}\caption{Illustration of the ion trap crystal with 100 ions. They are labeled according to the $z$ coordinates. }
    \label{figS3}
\end{figure}
The qubit state is encoded in the $|0\rangle\equiv|^{2}S_{1/2},F=0,m_F=0\rangle$ and $|1\rangle\equiv|^{2}S_{1/2},F=1,m_F=0\rangle$ levels of the ${}^{171}\mathrm{Yb}^+$ ions. 
The position of each ion in the $N=100$ ion crystal is shown in Fig.~\ref{figS3}. They are labeled in ascending order of their $z$ coordinates. 

\subsection{Ising model Hamiltonian}
\label{sec:Ising}
Two orthogonal pairs of counter-propagating 411 nm lasers generate a spin-dependent optical dipole force by imparting momentum to the ions via photon scattering. The spin state of each ion modulates its coupling to the laser field, creating a position-dependent force that displaces the ions' collective vibrational modes (phonons). These phonons act as a bosonic mediator: photons from the counter-propagating lasers coherently couple the spins to the phonon modes, while subsequent phonon exchange between ions propagates effective spin-spin interactions. Under the Lamb-Dicke confinement, the phonon dynamics occur on a timescale much faster than the spin-spin coupling rate. This separation of timescales allows adiabatic elimination of the phonons, leaving an effective Ising Hamiltonian with long-range coupling. The resulting interaction range and strength are tunable via the laser detuning, intensity, and ion crystal geometry. A full derivation is provided in the Supplementary Information of ref.~\cite{Guo2024Siteresolved}. In the end, we have:
\begin{align}
H_z = \sum_{ij} J_{ij} (I+\sigma_z^i) (I+\sigma_z^j)
=\sum_{i\ne j} J_{ij} \sigma_z^i \sigma_z^j + \sum_i h_i \sigma_z^i,
\label{eq:general}
\end{align}
where the lasers' detunings are $\Delta\pm \mu/2$ such that $|\Delta\pm \mu/2|\gg \Omega$ compared with the Rabi frequency of each laser beam. $\Omega^i_{\mathrm{eff}}\equiv\Omega_i^2/\Delta$ is the AC Stark shift on the ion $i$, $\omega_k$ is the frequency of the mode $k$, $\eta_k$ is the Lamb-Dicke parameter, and $b_{ik}$ is the mode vector's $i$th component. The coupling coefficients $J_{ij}$ are given by
\begin{equation}
J_{ij} = \frac{\Omega^i_{\mathrm{eff}} \Omega^j_{\mathrm{eff}}}{16}\sum_k \frac{\eta_k^2 b_{ik} b_{jk}}{\delta_k}, \label{eq:Jij}
\end{equation}
where $\delta_k=\mu-\omega_k$. The longitudinal field strength $h_i \equiv 2\sum_j J_{ij}$ is small and can be neglected: Assuming a nearly constant $\Omega^i_{\mathrm{eff}}$, we have $h_i\propto \sum_k b_{ik}$, which is equal to zero for all the modes apart from the COM mode. As for the COM mode, we can compensate this nearly constant longitudinal field by a small shift in the detuning of the two pairs of the $411\,$nm laser beams, so that a small asymmetry in their time-independent AC Stark shift can be generated. 

\subsection{Coupled to a single phonon mode}
When we tune the laser so that the spins mainly couple to a single mode $k$, the Hamiltonian can be simplified as 
\begin{equation}
H_z = \sum_{i\ne j} \frac{\Omega^i_{\mathrm{eff}} \Omega^j_{\mathrm{eff}}}{16} \frac{\eta_k^2 b_{ik} b_{jk}}{\delta_k}\sigma_z^i\sigma_z^j,
\end{equation}
which can be written in the following form up to some finite constant:
\begin{equation}
    H_z 
=\frac{1}{16\delta_k}(\sum_{i} \Omega^i_{\mathrm{eff}}\eta_k b_{ik} \sigma_z^i )^2.
\end{equation}
The Hamiltonian commutes with the single spin operator $\sigma_z^i$, so its eigenstates should have well-defined $\sigma_z^i$ for all ions. If we have negative $\delta_k$, the ground states are configured in a way that $\sigma_z^i=\pm \mathrm{sign}(b_{ik})$, where  $\pm$ comes from the $Z_2$ symmetry of the Hamiltonian. The correlation patterns of ground states in the $Z$ direction $\left\langle\sigma_z^i \sigma_z^j\right\rangle$ agree with that of the interaction matrix $J_{ij}$. 

We calculate the ground state and excited states, as well as their correlations with our method for the seventh phonon mode. To obtain the theoretical Ising model Hamiltonian, we first solve the equilibrium configurations of $N=100$ ions in a trap with trap frequencies $(\omega_x,\omega_y,\omega_z)=2\pi\times(0.690,2.140,0.167)\,$MHz. We sort the ions by their $z$ coordinates to match the ion indices used in the experiment. Then we solve the transverse phonon modes. We compute the theoretical $J_{ij}$ coefficients according to equation~\eqref{eq:Jij}. Specifically, we use a Rabi rate $\Omega_{\mathrm{eff}}=2\pi\times 10\,$kHz and a Lamb-Dicke parameter of $\eta\approx 0.11$ for the COM mode and counter-propagating $411\,$nm laser beams. We also add a transverse field in the $x$ direction $-h\sum_{i} \sigma_x^i$, which is applied in adiabatically preparing the ground state of the Ising model. The coupling strength $J_{ij} $ was originally expressed in kHz.
We nondimensionalized the system by rescaling time so that the kHz factor is absorbed, making $J_{ij} $ dimensionless.

\begin{figure}[t]
    \centering
    \includegraphics[width=\linewidth]{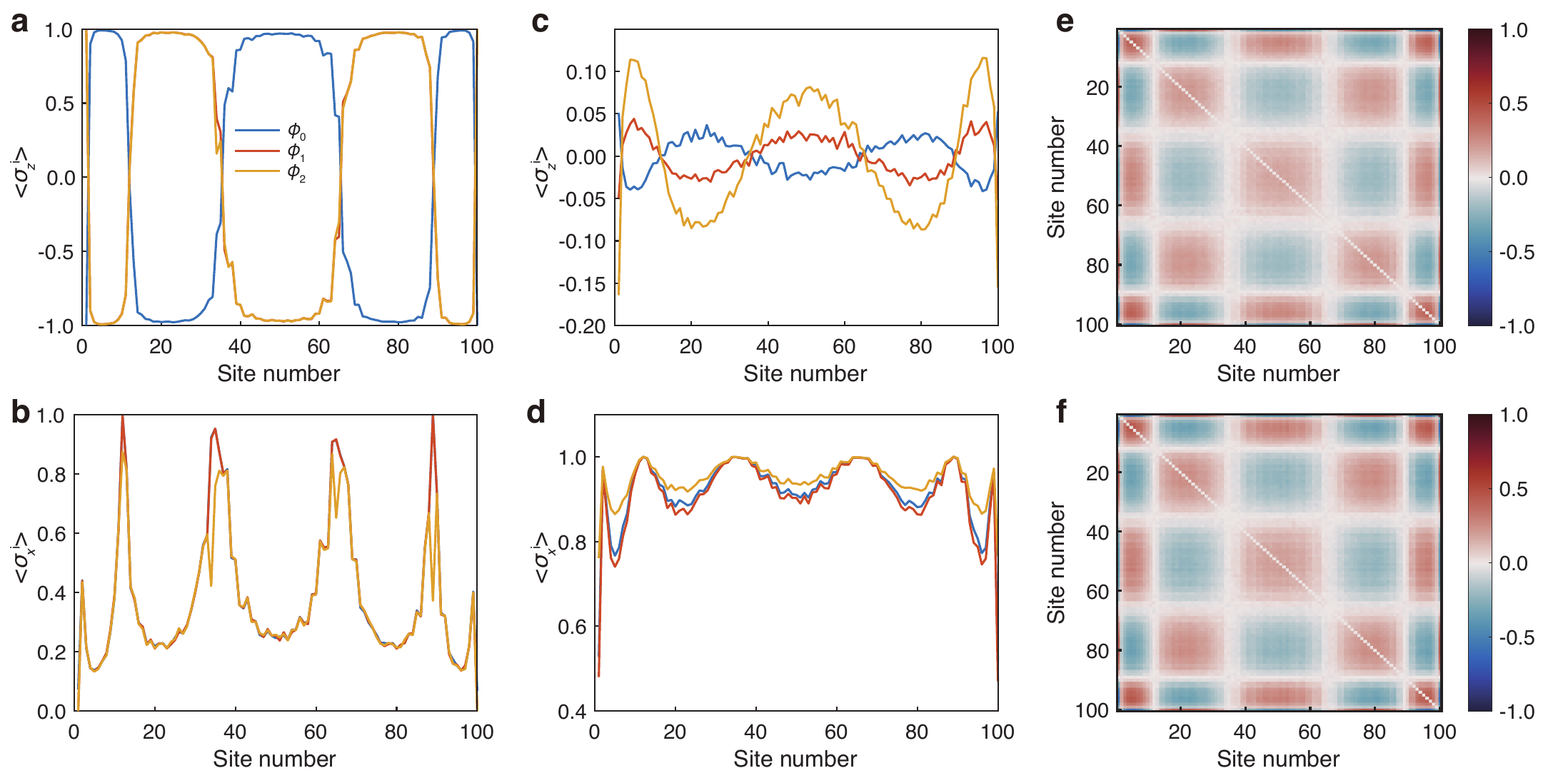}
    \caption{Extended NQES results for the 100 ions coupled to the seventh phonon mode. The transverse field $h$ is set to 2 for \textbf{a-b} and 8 for \textbf{c-f}. \textbf{a,c,} Expectation value of each spin in the $z$ direction $\left\langle\sigma_z^i \right\rangle, i=1,2,\cdots,100$ for the first three eigenstates $\phi_0,\phi_1,\phi_2$, respectively. \textbf{b,d,} Expectation value of each spin in the $x$ direction $\left\langle\sigma_x^i \right\rangle, i=1,2,\cdots,100$. \textbf{e-f,} Correlation maps $\left\langle\sigma_z^i\sigma_{z}^j\right\rangle,i,j=1,2,...,100$ for the ground state and the first excited state. The correlation map of the second excited state is shown in the main text. }
    \label{figS4}
\end{figure}

The Kac normalization of the coupling strength $J_{\mathrm{Kac}}=\sum_{ij} J_{ij}/N\approx 3.5$. In the main text, we show that the NQES can grasp the excited-state properties for different transverse field strengths. Here we present some extended results. In Fig.~\ref{figS4}\textbf{a-b}, we calculate the single-spin expectations $\langle \sigma_z\rangle $ and $\langle \sigma_x\rangle $ for each eigenstate with $h=2$. The sign of $\langle \sigma_z^i\rangle $ agrees with theoretical predictions, and its absolute value is close to 1. By contrast, the value of $\langle \sigma_x^i\rangle $ is small, except for certain site $i$ whose $b_{ik}$ is close to zero.  These results indicate a ferromagnetic phase.  In Fig.~\ref{figS4}\textbf{c-d}, we calculate the single-spin expectations $\langle \sigma_z\rangle $ and $\langle \sigma_x\rangle $ for each eigenstate with $h=8$. The value of $\langle \sigma_z^i\rangle $ appears small in this case, while $\langle \sigma_x^i\rangle $ is close to 1, indicating a paramagnetic phase. We also present the correlation maps for the ground state and the first excited state in Fig.~\ref{figS4}\textbf{e-f}. 

\subsection{Computing small systems with exact diagonalization}
\begin{figure}[t]
    \centering
    \includegraphics[width=1\linewidth]{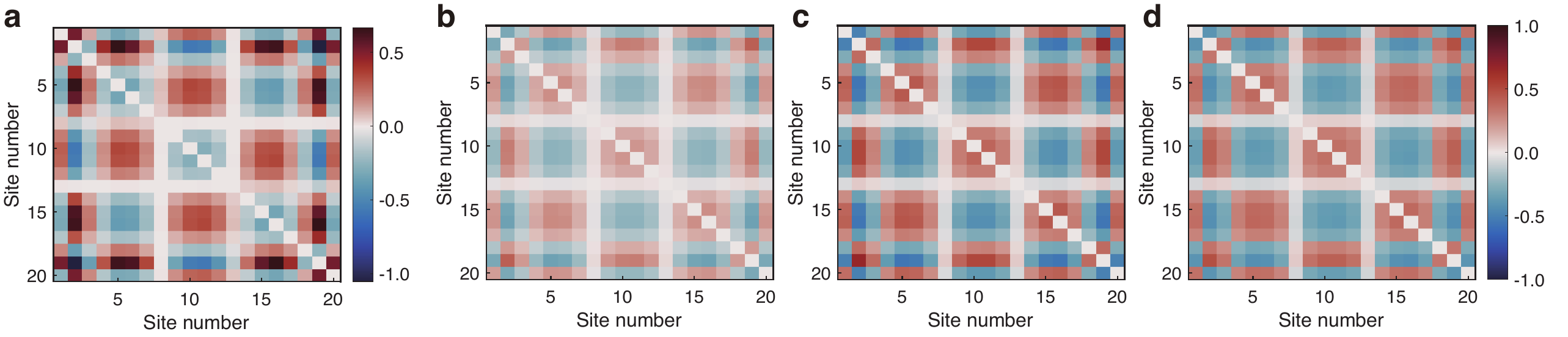}\caption{Exact diagonalization results for the 20 ions coupled to the seventh phonon mode. The transverse field $h$ is set to 8. \textbf{a,} Coupling matrix $J$. \textbf{b-d,} Correlation maps $\left\langle\sigma_z^i \sigma_{z}^j\right\rangle,i,j=1,2,...,20$ of the ground state, 1st excited state and 2nd excited state, respectively. }
    \label{figS5}
\end{figure}

\begin{figure}[t]
    \centering
    \includegraphics[width=1\linewidth]{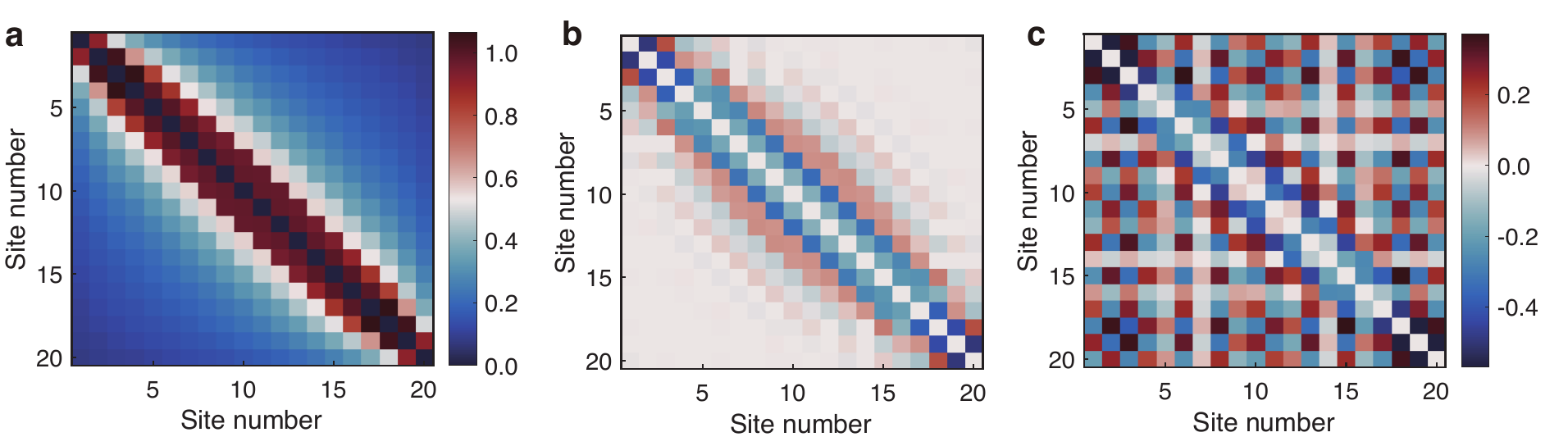}\caption{Exact diagonalization results for the 20 ions coupled to all phonon modes. The transverse field $h$ is set to 1.2. \textbf{a,} Coupling matrix $J$. \textbf{b-c,} Correlation maps $\left\langle\sigma_z^i \sigma_{z}^j\right\rangle,i,j=1,2,...,20$ of the ground state and 1st excited state, respectively. }
    \label{figS6}
\end{figure}

In Fig.~\ref{figS5} and Fig.~\ref{figS6}, we validate our results on small system sizes of 20 spins using exact diagonalization. We solved the first three eigenstates and obtained their correlation maps for ions coupled to the seventh phonon mode and all modes, respectively. 
We show that for ions coupled to a single mode, the correlation maps of low-lying excited states are quite similar in the presence of a relatively large transverse field $h=8$.  

For ions coupled to all phonon modes which feature power-law decaying long-range antiferromagnetic coupling, we observe that the ground state and excited states differ greatly on the correlation map with a transverse field $h=1.2$.  The ground state's correlation strength falls off with respect to distance, while the first excited state exhibits a long-range feature. 
All features have been observed in our calculation of hundreds of ions with NQES in the main text.

\section{Computational resources}

All simulations in this work were performed on four identical computational nodes, each equipped with dual AMD EPYC 7742 processors (64 cores per CPU, 128 cores per node) and 512 GB of RAM. Each calculation was executed independently on a single node, with no inter-node communication required.
\bibliography{QAI}